\documentclass[12pt]{article}

\usepackage[utf8]{inputenc}
\usepackage[english]{babel}
\usepackage{amsmath}	
\usepackage{amsfonts}
\usepackage{amssymb}
\usepackage{mathrsfs}
\usepackage[left=2cm,right=2cm,top=2cm,bottom=2cm]{geometry}
\usepackage[autostyle]{csquotes}
\usepackage[backend=biber,style=numeric-comp,sorting=none]{biblatex}
\usepackage{empheq}
\usepackage{braket}
\usepackage{tensor}
\usepackage[colorlinks,linkcolor=blue,citecolor=blue,urlcolor=blue]{hyperref}
\usepackage{breakurl}

\let\a=\alpha   \let\b=\beta   \let\g=\gamma   \let\d=\delta
\let\e=\epsilon    \let\h=\eta     \let\q=\theta
    \let\k=\kappa  \let\l=\lambda  \let\m=\mu
\let\n=\nu           \let\p=\pi      \let\r=\rho
\let\s=\sigma  \let\t=\tau      \let\f=\phi
     \let\y=\psi    \let\w=\omega
  \let\D=\Delta   \let\L=\Lambda
     \let\P=\Pi    \let\S=\Sigma  
\let\F=\Phi           \let\W=\Omega
\newcommand{\<}{\langle}
\renewcommand{\>}{\rangle}
\newcommand{\beq}{\begin{equation}}
\newcommand{\eeq}{\end{equation}}
\newcommand{\beqn}{\begin{eqnarray}}
\newcommand{\eeqn}{\end{eqnarray}}

\newcommand{\idnty}{\hbox{1$\!\!$1}}

\newcommand{\half}{\frac{1}{2}}

\newcommand{\Tr}{{\rm Tr}}
\newcommand{\tr}{{\rm tr}}

\newcommand{\eps}{\epsilon}

\newcommand{\be}{\begin{equation}}
\newcommand{\ee}{\end{equation}}
\newcommand{\ba}{\begin{eqnarray}}
\newcommand{\ea}{\end{eqnarray}}
\newcommand{\bdm}{\begin{displaymath}}
\newcommand{\edm}{\end{displaymath}}

\def\S{\Sigma}

\def\b{\beta}
\def\h{\eta}
\def\a{\alpha}
\def\g{\gamma}
\def\s{\sigma}
\def\k{\kappa}

\def\d{\delta}
\def\D{\Delta}

\newcommand{\im}{{\rm Im\,}}

\newcommand{\ie}{{\it i.e.\ }}
\newcommand{\eg}{{\it e.g.\ }}

\newcommand{\calA}{{\mathcal A}}
\newcommand{\calB}{{\mathcal B}}
\newcommand{\calC}{{\mathcal C}}

\newcommand{\calE}{{\mathcal E}}
\newcommand{\calF}{{\mathcal F}}
\newcommand{\calG}{{\mathcal G}}
\newcommand{\calH}{{\mathcal H}}
\newcommand{\calI}{{\mathcal I}}

\newcommand{\calM}{{\mathcal M}}
\newcommand{\calN}{{\mathcal N}}
\newcommand{\calO}{{\mathcal O}}
\newcommand{\calP}{{\mathcal P}}

\newcommand{\calR}{{\mathcal R}}

\newcommand{\calT}{{\mathcal T}}

\newcommand{\calW}{{\mathcal W}}
\newcommand{\calX}{{\mathcal X}}
\newcommand{\calY}{{\mathcal Y}}

\DeclareMathAlphabet{\mathpzc}{OT1}{pzc}{m}{it}
\def\bea{\begin{eqnarray}}
\def\eea{\end{eqnarray}}
\def\beas{\begin{eqnarray*}}
\def\eeas{\end{eqnarray*}}
\def\sla{\raise.15ex\hbox{$/$}\kern-.57em}

\def\bea{\begin{eqnarray}}
\def\eea{\end{eqnarray}}

\def\tr{{\rm tr}}
\def\sla{\raise.15ex\hbox{$/$}\kern-.57em}
\def\ie{{\it i.e.}~}
\def\eg{{\it e.g.}~}
\def\ap{{\alpha^\prime}}

\def\a{\alpha}
\def\b{\beta}
\def\g{\gamma}
\def\d{\delta}
\def\e{\epsilon}

\def\k{\kappa}
\def\l{\lambda}
\def\m{\mu}
\def\n{\nu}
\def\r{\rho}
\def\s{\sigma}
\def\t{\tau}
\def\f{\phi}

\def\w{\omega}

\def\F{\Phi}

\def\W{\Omega}
\def\cA{{\cal A}}

\def\cC{{\cal C}}

\def\cE{{\cal E}}
\def\cF{{\cal F}}
\def\cG{{\cal G}}

\def\cK{{\cal K}}

\def\cM{{\cal M}}
\def\cN{{\cal N}}
\def\cO{{\cal O}}
\def\cP{{\cal P}}
\def\cQ{{\cal Q}}
\def\cR{{\cal R}}
\def\cS{{\cal S}}

\def\cY{{\cal Y}}

\newcommand{\der}{{\partial}}											% Substitute \partial
\newcommand{\pder}[2]{\dfrac{{\partial}{#1}}{{\partial}{#2}}} 		% Partial derivative
\newcommand{\setZ}{{\mathbb{Z}}}
\newcommand{\setR}{{\mathbb{R}}}
\newcommand{\setC}{{\mathbb{C}}}

\let\e=\varepsilon							% Redefinition of epsilon
\newcommand{\levi}{{\epsilon}}					% Levi-Civita symbol

\newcommand{\partition}[1][]{\mathcal{Z}_\alpha^{{\mathcal{N}}{#1}}}
\newcommand{\even}[1][]{\mathbb{E}^{({#1})}}
\newcommand{\odd}[1][]{\mathbb{O}^{({#1})}}
\newcommand{\dbraket}[1]{\braket{\braket{#1}}}

\renewcommand{\url}{\burl} 

\makeindex		
\numberwithin{equation}{section}

\begin{document}
\begin{titlepage}
\begin{flushright}
{ROM2F/2015/10}\\
\end{flushright}
%
%\vskip 1cm
%
\begin{center}
{\Large\bf Simplifying one-loop amplitudes in superstring theory}
\end{center}
%\vskip 2cm
\begin{center}
%%%%%%%%%%%%%%%%%%%%%%%%%  AUTORI  %%%%%%%%%%%%%%%%%%%%%%%%%%%%%%%%%%%%%%%%%
{\bf Massimo Bianchi, Dario Consoli}\\
{\sl Dipartimento di Fisica, Universit\`a di Roma ``Tor Vergata''\\
 I.N.F.N. Sezione di Roma ``Tor Vergata''\\
Via della Ricerca Scientifica, 00133 Roma, Italy}
\end{center}
\vskip 1.0cm
%%%%%%%%%%%%%%%%%%%%%%%%%  ABSTRACT  %%%%%%%%%%%%%%%%%%%%%%%%%%%%%%%%%%%%%%
\begin{center}
{\large \bf Abstract}
\end{center} 
We show that 4-point vector boson one-loop amplitudes, computed in \cite{Bianchi:2006nf} in the RNS formalism, around  vacuum configurations with open unoriented strings, preserving at least $\cN =1$ SUSY in $D=4$, satisfy the correct supersymmetry Ward identities, in that they vanish for non MHV configurations $({+}{+}{+}{+})$ and $({-}{+}{+}{+})$. In the MHV case $({-}{-}{+}{+})$ we drastically simplify their expressions. 
We then study factorisation and the limiting IR and UV behaviours and find some unexpected results.
In particular no massless poles are exposed at generic values of the modular parameter.
Relying on the supersymmetric properties of our bosonic amplitudes, we extend them to manifestly supersymmetric super-amplitudes and compare our results with those obtained in the $D=4$ hybrid formalism, pointing out difficulties in reconciling the two approaches for contributions from ${\cN =1,2}$ sectors.

%PACS: 11.25.-w, 11.25.Db,  11.25.Hf, 11.25.Mj, 11.25.Uv, 11.25.Wx

%keywords: Strings, 

\vfill

\end{titlepage}

%%%%%%%%%%%%%%%%%%%%%%%%%%%%%%%%%%%%%%%%%%%%%%%%%%%%%%%%%%%%%%%%%%%%%%%%%%%%%%%%%%%%%%%%%%%%%%%%%%

\tableofcontents

\section*{Introduction}
There is a revival of interest in superstring loop amplitudes from different perspectives \cite{Bianchi:2006nf, Matone:2005vm, Tourkine:2013rda, Witten:2013tpa,Witten:2013cia,Witten:2012bh,Green:2013bza}. 
The original one-loop computations by Brink, Green and Schwarz \cite{Green:1982sw} paved the way to remarkable developments both in string theory and in field theory related contexts. 
Some time ago the BGS amplitude for four vector bosons in $D=10$ Type I superstrings was generalised to $D=4$ in vacuum configurations with open and unoriented strings preserving at least $\cN =1$ supersymmetry \cite{Bianchi:2006nf}. Although the final result could be expressed in a compact form as a sum over the various sectors, only the contribution of the $\cN =4$ sector and the `irreducible' contributions of the  $\cN =1,2$ sectors could be easily seen to be proportional to the tree-level amplitude. Supersymmetry Ward identities imply that the only non-vanishing 4-point amplitudes be Maximally Helicity Violating (MHV) \cite{Brandhuber:2007vm}. Thus loop corrections should reproduce similar structures to tree-level amplitudes in supersymmetric vacuum configurations in $D=4$.

Aim of the present paper is to simplify the results of \cite{Bianchi:2006nf} using the spinor helicity formalism and to analyse the singularities of the resulting amplitudes. We will also comment on the soft behaviour of the amplitudes and compare our results with the ones obtained in \cite{Berkovits:2001nv} within the $D=4$ `hybrid formalism' \cite{Berkovits:1994wr}\footnote{Other hybrid approaches have been proposed in diverse dimensions: $D=6$ \cite{Berkovits:1999im}, $D=3$ \cite{Berkovits:1999in}, $D=2$ \cite{Berkovits:2001tg}.} and find it hard to reconcile the two approaches. The main difference is that despite vector boson vertex operators are compactification independent, \ie they are proportional to the identity operator of the internal SCFT, only in the  
$\cN =4$ sector one has a complete factorisation of the space-time and internal part, encoded in the sum of KK momenta or alike. In $\cN =1,2$ sectors the computation does not factorise even for the `irreducible' contributions, encoded in a function $\cF_\calN$ of the modular parameter and of the compactification moduli, which does not simply coincide with the internal partition function. 

The plan of the paper is as follows. 
In Section \ref{sect:superstrings_at_one_loop}, we will briefly review one-loop open superstring amplitudes in order to fix the notation. In Section \ref{sect:two_point_amplitudes} and \ref{sect:three_point_amplitudes} we rewrite 2- and 3-point `amplitudes'\footnote{We put amplitudes in quote because they do not correspond to ``scattering'' due to collinear momenta.}, which vanish in the case of $\cN =4$ sector, in the helicity formalism. We show that they satisfy the correct Ward identities, in that, for instance, 3-point amplitudes with 3 positive helicity vector bosons vanish, and identify their divergences. The results of \cite{Bianchi:2006nf} for 4-point amplitudes are reviewed in Section \ref{sect:four_point_amplitudes} and systematically simplified in Section \ref{sect:simplifying_four_pt_amplitude}, where we show that only MHV amplitudes are non-vanishing. The case of `regular' branes at orbifold singularities \cite{Kachru:1998ys, Zoubos:2010kh} is discussed in Section \ref{sect:regular_branes}. In Section \ref{sect:UV_IR} we study factorisation and find some unexpected result, \ie no massless poles appear for generic modular parameter, even in sectors with reduced SUSY. We then discuss the IR and UV behaviours of the independent amplitudes: planar, non-planar ($3{-}1$ and $2{-}2$) and unorientable. After extending our bosonic amplitudes to full super-amplitudes, in Section \ref{sect:SUSY_vs_hybrid} we draw a comparison between our results and the results of the $D=4$ hybrid formalism \cite{Berkovits:2001nv} and we discuss potential sources of disagreement. 
We will conclude with some speculations about higher points, higher loops, soft limits and KLT relations beyond tree-level in Section \ref{sect:conclusions}.

\section{Superstrings at one loop}
\label{sect:superstrings_at_one_loop}

In theories with open and unoriented strings scattering amplitudes can be computed inserting the corresponding vertex operators on the boundaries \cite{Kiritsis:2007nutshell}. The vector boson vertex (in the super-ghost pictures $q=0$) reads
\begin{equation}
V^{(0)}_B= a_\mu \left(\partial X^\mu +i k{\cdot}\psi \psi^\mu \right) e^{ikX}= \left( a {\cdot}\partial X -\dfrac{i}{2} f_{\mu \nu}\psi^\mu \psi^\nu \right) e^{ikX}
\end{equation}
where $f_{\mu \nu}= a_\mu k_\nu - k_\mu a_\nu$ is the linearised field strength.\footnote{Henceforth we will refer to 
${:}\y_\m \y_\n{:}$ as (fermionic) bilinear.}

The tree-level disk contribution is similar to Veneziano amplitude
\begin{equation}
\cA_4^\textup{tree}(1,2,3,4) =  g^2_s F^4 \left[\dfrac{\calB(s,t)}{s t} \calT_{1234} + \dfrac{\calB(t,u)}{t u}\calT_{1423} + \dfrac{\calB(u,s)}{u s} \calT_{1342}\right]
\end{equation}
where $g_s$ is the coupling constant for open string. The Veneziano factor $\calB(x,y)$ and the Chan-Paton factor $\calT_{abcd}$ read
\begin{equation}
\calB(x,y) = \dfrac{\Gamma(1-\ap x) \Gamma(1-\ap y)}{\Gamma(1-\ap x-\ap y)} \quad \quad , \quad \quad
\calT_{abcd} = \tr(t_a t_b t_c t_d)
\end{equation}
while the totally symmetric kinematic factor $F^4$ is given by
\begin{equation}
F^4 = [2 (f_1f_2f_3f_4)-\frac{1}{2}(f_1f_2)(f_3f_4)  + {\rm cyclic(234)}]
\end{equation}
In $D=4$, $F^4$ is non-vanishing only in the Maximally Helicity Violating (MHV) case
\begin{equation}
F^4_{++++}=F^4_{-+++}=0 \quad \quad , \quad \quad F^4_{--++}= \dfrac{\<12\>^3}{\<23\>\<34\>\<41\>} st
\end{equation}
thereby, for a given color ordering, the partial amplitudes read
\begin{equation}
\cA_4^\textup{tree}[1^-,2^-,3^+,4^+] = \cA^\textup{tree}_\textup{YM}[1^-,2^-,3^+,4^+] \calB(s,t)
\end{equation}
while $\cA_4^\textup{tree}[1^+,2^+,3^+,4^+] = 0 = \cA_4^\textup{tree}[1^-,2^+,3^+,4^+]$. Generalization to higher points can be found in \cite{Mafra:2011nv,Mafra:2011nw,Barreiro:2012aw,Barreiro:2013dpa}.

The one-loop four-point amplitude in $D=10$ was computed long ago by Brink, Green and Schwarz \cite{Green:1982sw}. It receives three contributions: planar, non-planar and un-orientable. 
Setting the modular parameter of the covering torus to be $\tau_A = i{T}/2$ for the annulus and $\tau_M = i{T}/2 + 1/2$ for the M\"obius strip, all contributions can be written in the form
\begin{equation}
\cA^{\rm 1-loop}_4 (1,2,3,4)= g_s^4 \calT_{CP} \ap^2 F^4 \int_0^\infty \frac{dT}{T^{5+1}} \int_{\cR_{CP}} \!\!d^4 z_i\, \Pi_4(z_i,k_i)
\end{equation}
where $\calT_{CP}$ is the Chan-Paton factor, $\cR_{CP}$ is the integration region, depending on color ordering, and 
\begin{equation}
\Pi_4(z_i,k_i) = \prod_{i<j} \exp[-2\ap k_i {\cdot }k_j \cG(z_{ij})]
\end{equation}
is the ubiquitous Koba-Nielsen factor with $\cG(z_{ij})$ the scalar propagator (Bargmann kernel) for boundary insertions at one-loop 
\begin{equation}
\label{eq:scalar_propagator}
\calG_\calA(z_1,z_2;\t_\calA)=-\left[\log\dfrac{\q_1 (z_1-z_2|\t)}{\q_1^\prime (0|\t)}-2\p \dfrac{[\im(z_1-z_2)]^2}{\im \t} \right]
\end{equation}
We will often write $\calG_{ij}$ instead to $\calG(z_{ij})$. 

In the planar case all vertex operators are inserted on the same boundary of an annulus 
\begin{equation}
\calT^{\rm plan}_{1234} = \tr(t_1t_2t_3t_4) \tr(1) \qquad , \qquad \cR^{\rm plan}_{1234} = \{ z_1>z_2>z_3>z_4=0\}
\end{equation}
plus cyclic permutations of 234. The parametrization of the world-sheet variable on a boundary is $z=iT\n/2$ with $\n \in [0,1]$.

In the non-planar case vertex operators are equally distributed among the two boundaries of an annulus
\begin{equation}
\calT^{\rm non-pl}_{12|34} = \tr(t_1t_2) \tr(t_3t_4) \qquad , \qquad \cR^{\rm non-pl}_{12|34} = \{ z_1>z_2 ; z_3>z_4=0\}
\end{equation}
plus permutations of 2 with 3 and 4. The parametrization of the world-sheet variable on the other boundary is $z=iT\n/2+1/2$ with $\n \in [0,1]$.

For gauge groups with (anomalous) $U(1)$ factors there is an additional non-planar contribution with 3 vertices inserted on a boundary and the remaining one on the other boundary 
\begin{equation}
\calT^{\rm anom}_{123|4} = \tr(t_1t_2t_3) \tr(t_4) \qquad , \qquad \cR^{\rm anom}_{123|4} = \{ z_1>z_2>z_3=0; z_4\}
\end{equation}
plus permutations of 4 with 1,2,3. 

In the un-orientable case vertex operators are inserted along the single boundary of a M\"obius strip with twice the length of the strip itself
\begin{equation}
\calT^{\rm un-or}_{1234} = 2 \tr(t_1t_2t_3t_4) \calT_{\Omega} \qquad , \qquad \cR^{\rm un-or}_{1234} = \{ z_1>z_2>z_3>z_4=0\}
\end{equation}
plus cyclic permutations of 234, with $\calT_\Omega$ the tension of the relevant $\Omega$-plane in units of $\ap$, quantized charge. The parametrization of the world-sheet variable on the unique boundary is $z=iT\n/2$ with $\n \in [0,2]$.

At low-energies $\ap | k_i{\cdot}k_j | <<1$, $\Pi_4(z_i,k_i)\approx 1$ and one can trivially integrate over the insertion points $z_i$ producing a factor of ${T}^4$. The remaining integral over the modular parameter $\int d{T}/{T}^2$ is IR finite (for ${T}\rightarrow \infty$) but UV divergent (for ${T}\rightarrow 0$), due to the dilaton tadpole associated to the empty boundary and the `cross-cap'. Yet for $SO(32)$, the dilaton tadpole cancels and the Type I theory if free of both UV divergences and chiral anomalies \cite{Green:2012pqa}. Non-planar amplitudes are regulated by momentum flow between the two boundaries. 

A subtle issue related to potential anomalies is the role of the odd spin structure in the computation of scattering amplitudes. In order to detect potential anomalies, one of the gauge boson vertex operators should appear with longitudinal polarisation and should decouple thanks to BRST invariance in a consistent theory. The standard procedure requires the insertion of a vertex  operator in the $q=-1$ super-ghost picture and an additional world-sheet super-current insertion brought down by integration over the super-modulus associated to the world-sheet gravitino zero. In $D=10$ hexagon gauge anomaly could be detected this way \cite{Green:1984qs,Gross:1987pd}. In the $D=4$ case under consideration, 4-point amplitudes are not anomalous and BRST invariance allows to replace the combination of the super-modulus and world-sheet super-current with a picture changing operator \cite{Friedan:1985ey}. The latter can act on the vertex  operator in the $q=-1$ super-ghost picture and change its picture $q=0$.  As in \cite{Bianchi:2006nf}, one can then proceed with all vertex operators in the $q=0$ picture\footnote{This argument must be taken with a grain of salt since it has caused a problematic impression on the referee. According to the referee, that we thank for his/her punctual observation, it leads to an incorrect number of fermionic propagators, \ie $S_{ij}^{n-2}$ instead of $S_{ij}^{n-1}$. The latter is what one might expect in line with the counting of loop momenta in the field theory limit. Yet, our treatment of the odd spin structure precisely matches the results in the even spin structures in so far as the counting of $S_{ij}$ is concerned. Moreover, we checked that our procedure reproduces the results of \cite{Gross:1987pd} since the longitudinal vertex in the $q=0$ picture is a total derivative that leads to a vanishing result in a consistent theory or to a boundary term as a signal of an anomaly. We plan to return to this issue in the future.}.

\subsection{Partition function}

In order to generalize BGS formula to (supersymmetric) vacuum configurations for open and unoriented strings in $D=4$, one has to first recall the structure of the one-loop partition function. As in \cite{Bianchi:2006nf}, we will mainly focus here on magnetised or intersecting D-branes at (non-compact) orbifold singularities \cite{Kachru:1998ys, Zoubos:2010kh}. Consistency requires local RR tadpole cancellation \cite{Bianchi:1988fr,Bianchi:1990tb,Bianchi:1990yu,Angelantonj:1996uy,Angelantonj:2002ct}.

The partition function depends on the choice of brane configuration, including number and type of intersections or magnetic fluxes thereon, $\Omega$-planes and orbifold group $\Gamma$. For simplicity we will focus on $\Gamma=\setZ_n \subset SU(3)$, \ie $Z^I \approx \exp(2\pi i n_I/n) Z^I$ with $n_1+n_2+n_3 = 0 ~({\rm mod}\, 1)$ in order to preserve SUSY. 
We will label the branes of type $a$ by $i_a=1,...,N_a$, and the orbifold sectors by ${{h}}=0,...,n-1$. We define three combinations that express the twist or shift in the open string spectrum 
\begin{equation}
u_{ab}^I= {{h}} v_{ab}^I + i \eps_{ab}^I \frac{T}{2} 
\end{equation}
where $I=1,2,3$, $\eps_{ab}^I$ denotes the angles between brane $a$ and brane $b$ or the shift in the string spectrum induced by the relative magnetic flux and $v_{ab}^I$ denotes the twist caused by the orbifold. The combinations $u_{ab}^I$ determine the amount of supersymmetry preserved in each sector.

\subsection{$\calN=1$ sectors}

The weakest condition one can impose to have `minimal', {\it i.e.~} $\calN=1$, supersymmetry is
$u^1_{ab}+u^2_{ab}+u^3_{ab} =0$, with $\textstyle \prod_I u^I_{ab}\neq 0$.
In this case the partition function assumes the form
\begin{equation}
\partition[=1] =\calX_{ab}^{\calN=1} \dfrac{\theta_\a(0)}{\h^3 } \prod_{I=1}^3 \dfrac{\q_\a (u_{ab}^I)}{\q_1 (u_{ab}^I)} \qquad {\rm with} \qquad  \calX_{ab}^{\calN=1}= \dfrac{ {\cal V}_X I_{ab}}{2_\textup{GSO}2_\W n_\textup{orb} (\ap T)^2}
\end{equation}
where $\a$ labels the four spin structures, ${\cal V}_X$ represents the `regulated' volume of space-time (to be replaced by $(2\pi)^4\delta(\Sigma_i k_i)$ in scattering amplitudes), $I_{ab}$ denotes the number of brane intersections or the degeneracy of Landau levels. We have traced the origin of various integers in the denominator, wherein the factor 
$(\ap T)^2$ accounts for integration over loop-momenta in $D=4$.

\subsection{$\calN=2$ sectors}
In sectors with $\calN=2$ supersymmetry one of the $u^I_{ab}$ vanishes, let us say $u^3_{ab}=0$.  As a consequence $u^2_{ab}=-u^1_{ab}$. The partition function reads
\begin{equation}
\partition[=2] =\calX_{ab}^{\calN=2} \dfrac{\q^2_\a(0) \q_\a^2(u_{ab}) }{\h^6 \q_1^2 (u_{ab})}
\qquad {\rm with} \qquad \calX_{ab}^{\calN=2}= \dfrac{ {\cal V}_X \L^{\parallel}_{ab} I^\perp_{ab}}{2_\textup{GSO}2_\W n_\textup{orb} (\ap T)^2}
\end{equation}
Where $u_{ab}=u^1_{ab}$, $I^\perp_{ab}$ denotes the number of intersections or degeneracy of Landau levels in the `twisted/transverse' directions and $\L^{\parallel}_{ab}$ denotes the lattice sum in the two (one complex) `untwisted/longitudinal' compact directions.

\subsection{$\calN=4$ sectors}
Sectors with $\calN=4$ maximal supersymmetry correspond to $u_{ab}^I=0$ and the partition function is simply given by
\begin{equation}
\partition[=4]= \calX_{ab}^{\calN=4} \dfrac{\q_\a^4(0)}{\h^{12}} \qquad {\rm with} \qquad
\calX_{ab}^{\calN=4}= \dfrac{{\cal V}_X \L_{ab} }{2_\textup{GSO}2_\W n_\textup{orb} (\ap T)^2}
\end{equation}
where $\L_{ab}$ denotes the lattice sum in the six internal directions. 

Later on, we will compute 2, 3 and 4-point scattering amplitudes. Although the first two formally vanish on-shell due to collinear kinematics, we report their derivation using the spinor helicity formalism since it highlights the meaning of some of the structures that will later appear in the more interesting 4-pt amplitudes. Definitions and notation for elliptic functions and helicity spinors can be found in the appendices. 

\section{Two-point amplitudes}
\label{sect:two_point_amplitudes}

Let us begin with the two-point amplitude without specifying for the time being whether the amplitude is planar, non-planar or un-oriented. We will see that the results are substantially the same up to minor modifications. Although momentum conservation implies $k_1{\cdot}k_2 =0$ and then $k_1{\cdot}a_2 =0 = k_2{\cdot}a_1$, we will formally keep $f_1f_2 \neq 0$. The one-loop amplitude is given by
\begin{equation}
\begin{split}
\calA_2^\textup{1-loop} [1,2]&= g_s^2 \sum_\a c_\a \int_0^\infty \! \dfrac{dT}{T} \! \int_0^{iT/2} \!\!\!\!\!\!dz_1 \int_0^{z_1} \!\!dz_2 \d(z_2) \braket{V^{(0)}_B (z_1)\,V^{(0)}_B (z_2)}_\a= \\
&=g_s^2 \int_0^\infty \dfrac{dT}{T} \int d\m^{(2)} (\even[2]+\odd[2])
\end{split}
\end{equation}
where $\even[2]$ and $\odd[2]$ denote the contributions of the even and odd spin structures. Contractions with zero and one bilinear are zero, we have only the two bilinears contribution. In the even spin structures, the reduced contraction of two bilinears yields
\begin{equation}
\even[2]=\sum_\a c_\a \dbraket{V^{(0)}_B (z_1)\,V^{(0)}_B (z_2)}_\a=-\dbraket{{:}k_1 {\cdot} \y_1\, a_1{\cdot}\y_1 {:} \, {:}k_2{\cdot}\y_2 \, a_2{\cdot}\y_2{:}}_\a =-\ap^2 \dfrac{(f_1 f_2)}{2} S_\a^2(z_{12})
\end{equation}
where $S_\a(z_{12})$ is the one-loop fermionic propagator (Szego kernel)\footnote{In the space orthogonal to the constant zero-mode.}
\begin{gather}
S_\a(z_1,z_2;\t)=
\begin{cases}
-\der_{z_1} \calG(z_1,z_2), & \text{if $\a=1$, \: {\rm odd}} \\
\dfrac{\q_\a(z_1-z_2)}{\q_1(z_1-z_2)} \dfrac{\q_1^\prime(0)}{\q_\a  (0)} , & \text{if $\a \neq 1$ \: {\rm even}.}
\end{cases}
\end{gather}
where $\calG$ is the scalar propagator in \ref{eq:scalar_propagator}. We often use the notation $S_{ij}$ instead of $S_1(z_{ij})=-\der_i \calG(z_{ij})$.
Using the identity $S_\a^2=\calP{-}e_{\a-1}$, where $\calP$ is Weierstrass function, that does not contribute to the sum over spin structures, and $e_{\a-1} = 2\pi i \der_\t \log(\q_\a/\h)$, we have
\begin{equation}
\even[2] = -\dfrac{\ap^2}{2} (f_1 f_2) \calE_\calN e^{-\ap k_1 {\cdot}k_2 \calG_{12}}
\end{equation}
\begin{equation}
\label{eq:E_definition}
\calE_\calN= -\sum_\a c_\a e_{\a-1}\partition
\end{equation}
where the function $\calE_\calN$ introduced in \cite{Bianchi:2006nf}, labelled by the number $\calN$ of preserved SUSY, depends on the world-sheet modular parameter $T$, on the parameters of the brane configuration coded in $u^I_{ab}$ and the moduli of the compactification. $\calE_\calN$ vanishes in $\calN=4$ sectors due to Riemann identity.

In the odd spin structure, which only contributes in $\calN=1$ sectors\footnote{In $\calN=2,4$ sectors one cannot absorb the internal fermionic zero-modes $\y_0^i$ with vector boson vertex operators.} we can absorb the four zero modes in a unique way
\begin{equation}
\odd[2]=-c_1^\textup{GSO}\braket{{:}k_1{\cdot} \y_0\, a_1{\cdot} \y_0{:}\,{:}k_2{\cdot}\y_0\, a_2{\cdot} \y_0{:}}= -2 \ap^2 c_1^\textup{GSO} \p^2 \calX_{\calN=1} (\tilde{f}_1 f_2)/T^2
\end{equation}
For brevity we define the function 
\begin{equation}
\label{eq:CN1_definition}
\calC_{\calN =1} =-2 c_1^\textup{GSO}\p^2 \calX_\calN/T^2
\end{equation}
Combining the contributions of even and odd spin structures, the two-point amplitude thus reads
\begin{equation}
\calA_2^\textup{1-loop} [1,2]= g_s^2 \ap^2 \int_0^\infty \dfrac{dT}{T} \left[\calC_{\calN=1} (\tilde{f}_1 f_2) + \calE_\calN (f_1 f_2) \right] \int d\m^{(2)}\, e^{-\ap k_1{\cdot} k_2 \calG_{12}}
\end{equation}
Fixing the helicities and noting that $k_1{\cdot} k_2 =0$ we have two simple results. Choosing $({\pm}{\pm})$ and using $\tilde{f}^\pm=\pm i f^\pm$ we obtain
\begin{equation}
\label{eq:amplitude_2_p_1_l_pp}
\calA^\textup{1-loop}_2[1^\pm,2^\pm]= -\dfrac{g_s^2 \ap^2}{4} (f_1^\pm f_2^\pm) \int_0^\infty \dfrac{dT}{T} \left[\calE_\calN \pm i \calC_{\calN=1} \right] T^2
\end{equation}
while choosing $({\pm}{\mp})$ the amplitude vanishes, {\it viz.} 
\begin{equation}
\label{eq:amplitude_2_p_1_l_pm}
\calA^\textup{1-loop}_2[1^\pm,2^\mp] = 0
\end{equation}
To obtain planar, non-planar and un-oriented contributions one has to choose the specific modular parameter and the corresponding integration domain. The result is generically divergent for both $\calN = 1,2$ sectors, wherein it encodes $\b$-functions and one-loop threshold corrections to the gauge kinetic functions \cite{Kiritsis:1997hf, Anastasopoulos:2006hn}. As already observed, it vanishes in $\calN = 4$ sectors, which points to the no-bubble conjecture in $\calN = 4$ SYM, \ie to the absence of one-loop massless amplitudes with two (bunches of) insertion points. 

\section{Three-point amplitudes}
\label{sect:three_point_amplitudes}
We continue our preliminary analysis and compute the three-point one-loop amplitude that reads:
\begin{equation}
\begin{split}
\calA_3^{\textup{1-loop}} [1,2,3]& = g_s^3 \sum_\a c_\a \int_0^\infty \dfrac{dT}{T} \int_0^{iT/2} \!\!\!\!\!\! dz_1 \int_0^{z_1} \!\!\!\! dz_2 \int_0^{z_2}\!\! dz_3\,\, \d(z_3) \braket{V^{(0)}_B (z_1)\,V^{(0)}_B (z_2)\,V^{(0)}_B (z_3)}_\a=\\
&=g_s^3 \int_0^\infty \dfrac{dT}{T} \int d\m^{(3)} \left(\even[3]+\odd[3]\right)
\end{split}
\end{equation}
Momentum conservation for massless vector bosons implies $k_i{\cdot}k_j = 0$ \ie collinear momenta. In order to proceed one could either relax momentum conservation \cite{Minahan:1987ha, Kiritsis:1997hf, Berg:2011ij, Berg:2014ama} yet with $(\sum_i k_i)^2 = 0$ or analytically continue to complex momenta \cite{Elvang:2013cua}. In the spinor helicity formalism, reviewed in appendix \ref{sect:spinor_helicity_formalism}, one has $2k_i{\cdot}k_j =- \langle ij\rangle [ij]$ and there are two options: either $\langle ij\rangle \neq 0$, with $[ij]=0$, convenient for MHV or $({-}{-}{-})$ helicity configurations or the other way around. 
In the even spin structures, we have two types of contributions from two and three bilinears. The term with three bilinears produces
\begin{equation}
\begin{split}
\even[3]_\textup{3-bil} &= -i \sum_\a c_\a \braket{{:}k_1 {\cdot} \y_1\, a_1 {\cdot} \y_1 {:} \, {:}k_2 {\cdot} \y_2 \, a_2 {\cdot} \y_2{:}\, {:}k_3 {\cdot} \y_3 \, a_3 {\cdot} \y_3{:}}_\a=\\
&=i\ap^3(f_1 f_2 f_3) \sum_\a c_\a S_\a(z_{12}) S_\a (z_{23}) S_\a (z_{13}) \partition \P_3(z_i,k_i) =-i \ap^3 (f_1 f_2 f_3) \w_{123} \P_3(z_i,k_i)
\end{split}
\end{equation}
where 
\begin{equation}
\label{eq:defomega123}
\w_{123}=S_{12}+S_{23}+S_{31} \quad ,
\end{equation}
$\Pi_3(z_i,k_i)=\textstyle{\prod_{i<j} e^{- \ap k_i{\cdot} k_j \calG_{ij}}}$ is the Koba-Nielsen factor and in the last step we have used 
\begin{equation}
S_\a(z_{12})S_\a(z_{23})=-\w_{123}S_\a (z_{13})-S^\prime_\a (z_{13}) \quad 2 S_\a(z_{12}) S_\a^\prime(z_{12})=\calP^\prime_{12}
\end{equation}
The terms with two bilinears produce
\begin{equation}
\even[3]_\textup{2-bil}=-\sum_\textup{cyclic}\sum_\a c_\a \braket{{:}a_1 {\cdot} \der X_1 {:}\, {:}k_2{\cdot} \y_2 \, a_2{\cdot} \y_2{:}\, {:}k_3{\cdot} \y_3 \, a_3{\cdot} \y_3{:}}_\a = i\dfrac{\ap^3}{2} \calE_\calN \sum_\textup{cyclic} a_3 {\cdot}P_3 (f_1 f_2) \P_3(z_i,k_i)
\end{equation}
where $P_i^\m=\textstyle{\sum_{j\neq i} k^\m_j S_{ij}}$, with $S_{ij} = -\partial_i \calG_{ij}$.

In the odd spin structure we have similar contributions, from three bilinears we have three terms depending on the choice of the points where the two zero modes are absorbed\footnote{Absorbing all the zero modes at two points would give zero due to normal ordering at the remaining point.}
\begin{equation}
\odd[3]_\textup{3-bil}{=}{-}i c_1^\textup{GSO} \!\!\sum_\textup{swaps} \!\!\braket{{:}k_1{\cdot}\y_0\, a_1{\cdot}\y_0{:}\,{:}k_2{\cdot}\y_0\, a_2{\cdot} \y{:}\,{:} k_3{\cdot}\y_0\, a_3{\cdot}\y{:}}{=}{-}i\ap^3 \calC_{\calN=1} \!\!\sum_\textup{cyclical} \!\!(\tilde{f}_1 f_2 f_3) S_{23} \P_3(z_i,k_i)
\end{equation}
where the sum on exchanges means summing terms with $\y$ and $\y_0$ exchanged in the same vertex. From two bilinears we have three terms
\begin{equation}
\odd[3]_\textup{2-bil}=-\sum_\textup{cyclical}\braket{{:}a_1{\cdot}\der X{:}\,{:}k_2{\cdot}\y_0\, a_2{\cdot}\y_0{:}\,{:}k_3{\cdot}\y_0\, a_3{\cdot} \y_0{:}}=\dfrac{i}{2} \ap^3 \calC_{\calN=1} \sum_\textup{cyclical} a_3 {\cdot}P_3 (\tilde{f}_1 f_2) \P_3(z_i,k_i)
\end{equation}

Let us consider the two independent helicity configurations. First the case $({+}{+}{+})$. We begin from even spin structures, we must compute the scalar products $a_i {\cdot}P_i$:
\begin{equation}
a_3^+ {\cdot} P_3 (f_1^+ f_2^+) {=} {-}\dfrac{[12]^2}{\sqrt{2}} \left(\dfrac{[31]\<1q\>}{\<3q\>} S_{31}{+}\dfrac{[32]\<2q\>}{\<3q\>}S_{32} \right)
{=}(f_1^+ f_2^+ f_3^+) \left(\dfrac{[12]\<1q\>}{[23]\<3q\>} S_{31}{-}\dfrac{[12]\<2q\>}{[31]\<3q\>}S_{32} \right)
\end{equation}
Using momentum conservation $[12]\<1q\>=-[32]\<3q\>$ and $[12]\<2q\>=-[13]\<3q\>$, so that one has $a_3^+ {\cdot} P_3 (f_1^+ f_2^+) =$ $(f_1^+ f_2^+ f_3^+) (S_{31}+S_{23})$ and one can thus factor out $(f_1^+ f_2^+ f_3^+)$ and get
\begin{equation}
\even[3]=\dfrac{i}{2}\ap^3 \calE_\calN (f_1^+ f_2^+ f_3^+) \sum_\textup{cyclic}  ( S_{31} - S_{23}) \P_3(z_i,k_i)=0
\end{equation}
The contribution of the odd spin structure becomes proportional to the contribution of the even spin structure after using $\tilde{f}=if$, thus the complete amplitude vanishes for this choice of helicities, as well as for $({-}{-}{-})$,
\begin{equation}
\label{eq:amplitude_3_p_1_l_ppp}
\calA_3^\textup{1-loop}[1^+,2^+,3^+] = 0 = \calA_3^\textup{1-loop}[1^-,2^-,3^-]
\end{equation}
as expected from SUSY Ward identities\footnote{We thank Nathan Berkovits for raising this issue.}. Indeed, barring anomalous $U(1)$'s, the only supersymmetric invariant for 2- and 3-points is $W^\alpha W_\alpha$ giving rise to the standard $F^2$ bosonic term, since $F^3$, though present in the bosonic string even at tree level, does not admit a SUSY completion.

Let us then consider the case $({-}{+}{+})$. For this helicity configuration one has a single term. The even spin structures produce
\begin{equation}
\even[3] = \dfrac{i}{2} \ap^3 \calE_\calN a_1^- {\cdot} P_1 (f_2^+ f_3^+) \P_3(z_i,k_i)
\end{equation}
It is convenient to compute $a_1^-{\cdot} P_1 (f_2^+ f_3^+)$ factorizing the MHV amplitude. In $a_1^-{\cdot} k_i$ the factor $\<1i\>$ would give zero due to collinear kinematics. In order to circumvent these subtleties, one can analytically continue to complex momenta and choose $q_1=k_3$. This yields the `right' result:
\begin{equation}
a_1^-{\cdot} P_1 (f_2^+ f_3^+)= -\dfrac{1}{\sqrt{2}} \dfrac{\<12\>[23]}{[13]} [23]^2 S_{12}=
- \sqrt{2} \dfrac{[23]^3}{[12][31]} k_1 {\cdot} k_2 S_{12}
\end{equation}
Using partial integration one can replace $k_1{\cdot} k_2 S_{12}$ with $ k_2 {\cdot}k_3 S_{23}$ to make it look more symmetric and finally find
\begin{equation}
\even[3]= -\dfrac{i}{\sqrt{2}} \ap^3 \calE_\calN \dfrac{[23]^3}{[12][31]} k_2 {\cdot}k_3 S_{23} \P_3(z_i,k_i)
\end{equation}
In the odd spin structure, only $\calN = 1$ sectors contribute. Following similar steps, one finds a similar result with $\calE_\calN$ replaced by $-i \calC_\calN$. Now we can write the complete three-point amplitude
\begin{equation}
\label{eq:amplitude_3_p_1_l_mpp}
\calA_3^\textup{1-loop}[1^-,2^+,3^+]= -\dfrac{i}{\sqrt{2}}g_s^3 \ap^3 \dfrac{[23]^3}{[12][31]} \int_0^\infty \dfrac{dT}{T} \int d\m^{(3)} \left(\calE_\calN-i\calC_{\calN=1} \right) k_2{\cdot} k_3 S_{23} \P_3(z_i,k_i)
\end{equation}

To compute planar, non-planar and un-oriented contributions one has to choose the specific modular parameter and the corresponding integration domain \cite{Sagnotti:1987tw, Bianchi:1988fr, Pradisi:1988xd, Bianchi:1988ux, Bianchi:1989du}. The result is generically divergent for both $\calN = 1,2$ sectors, wherein it is related by gauge invariance to the 2-point amplitude, encoding $\b$-functions and one-loop threshold corrections to the gauge kinetic functions. As already observed, it vanishes in $\calN = 4$ sectors, in a way reminiscent of the no-triangle conjecture in $\calN = 4$ SYM, \ie the absence of one-loop massless amplitudes with three (bunches of) insertion points.

\section{Four-point amplitudes}
\label{sect:four_point_amplitudes}

We are now ready to compute four-point amplitudes. We start by briefly reviewing and summarising the results of \cite{Bianchi:2006nf} and then analyse them  
in terms of helicity configurations, color orderings and limiting behaviours.

The starting point is
\begin{equation}
\begin{split}
\calA_4^\textup{1-loop}[1,2,3,4] &= g_s^4 \sum_\a c_\a \int_0^\infty \dfrac{dT}{T} \int d\m^{(4)} \braket{V_0 (1) V_0 (2) V_0 (3) V_0 (4)}_\a =\\
& = g_s^4 \int_0^\infty \dfrac{dT}{T} \int d\m^{(4)} \left(\even[4]+\odd[4]\right)
\end{split}
\end{equation}
where the integration region $\calR_{CP}$ and the Chan-Paton factor depend on the distributions of insertions on the two boundaries for the annulus (planar $4{-}0$, non-planar $3{-}1$ and $2{-}2$). For the un-orientable case there is no choice, except for the relative ordering of the insertions. Here, we will only summarise the results, the details can be found in \cite{Bianchi:2006nf}.
\subsection{Even spin structures, four bilinears}
\begin{equation}
\even[4]_\textup{4-bil}= \sum_\a c_\a \dbraket{{:}k_1 {\cdot} \y_1\, a_1 {\cdot} \y_1{:} \dots {:}k_4 {\cdot} \y_4\, a_4 {\cdot}\y_4{:}}_\a \partition
\end{equation}
The fermionic contribution consists in two types of terms connected and disconnected. The result for connected contractions is
\begin{equation}
\even[4]_\textup{4-bil,conn}=\ap^4 \sum_\textup{conn} (f_1 f_2 f_3 f_4) \left[\dfrac{1}{2}\calE_\calN(\calP_{13}+\w_{123}\w_{341}+\calP_{24}+\w_{234}\w_{412})-\calF_\calN \right] \P_4(z_i,k_i)
\end{equation}
where $\cE_{\cN}$ (vanishing for $\cN =4$) was defined previously in (\ref{eq:E_definition}) and $\w_{ijk}$ are defined in \ref{eq:defomega123}, while 
\begin{equation}
\label{eq:F_definition}
\calF_\calN=\sum_{\alpha=2}^4 c_\alpha e_{\alpha-1}^2 \partition
\end{equation}
depends on the number $\calN$ of preserved SUSY, on the world-sheet modular parameter $T$, on the parameters of the brane configuration coded in $u^I_{ab}$ and the moduli of the ``compactification". The disconnected contractions yield
\begin{equation}
\even[4]_\textup{4-bil,disconn}= \ap^4 \sum_\textup{disconn} \dfrac{1}{4} (f_1 f_2) (f_3 f_4) \left[\calE_\calN \left(\calP_{12}+\calP_{34}\right)+ \calF_\calN \right] \P_4(z_i,k_i)
\end{equation}

\subsection{Even spin structures, three bilinears}

Aside from the bosonic contractions, the fermionic contractions are the same as for three-point amplitudes, discussed previously, thus one finds
\begin{equation}
\begin{split}
\even[4]_\textup{3-bil}&= -i \sum_\textup{cyclic} \sum_\a c_\a \dbraket{{:}a_1 {\cdot} \der X_1 {:}\,{:} k_2 {\cdot} \y_2\, a_2 {\cdot}\y_2 \,k_2 {\cdot} \y_2 {:}\,{:} a_3 {\cdot}\y_3\, k_3 {\cdot} \y_3 {:}\,{:} a_4 {\cdot}\y_4 {:}}_\a \P_4(z_i,k_i) \partition=\\
&= -\ap^4 \sum_\textup{cyclic} a_1 P_1 (f_1 f_2 f_3) \w_{123} \calE_\calN \P_4 (z_i,k_i)
\end{split}
\end{equation}

\subsection{Even spin structures, two bilinears}

We already computed the fermionic contractions, thus the contribution to the amplitude is
\begin{equation}
\begin{split}
\even[4]_\textup{2-bil}=&-\sum_\textup{pairs}\dbraket{ {:}a_1 {\cdot} \der X_1{:}\,{:}a_2 {\cdot} \der X_2{:}} \dbraket{{:}k_3 {\cdot} \y_3\, a_3 {\cdot} \y_3 {:} \, {:}k_4 {\cdot} \y_4 \, a_4 {\cdot} \y_4{:}}_\a \partition=\\
&=-\dfrac{\ap^3}{2} \calE_\calN \sum_\textup{pairs} (f_3 f_4) \left(a_1{\cdot} a_2 \der_1 \der_2 \calG_{12} - \ap a_1 {\cdot}P_1 a_2{\cdot} P_2 \right) \P_4 (z_i,k_i)
\end{split}
\end{equation}
Notice that each term is gauge invariant {\it per se} up to total derivatives. For instance, replacing $a_1$ (or $a_2$) with the momentum $k_1$ and noting that $\der_{z_i} \P_4=-\ap k_i{\cdot} P_i \P_4$ (with $\P_4=\P_4(z_i,k_i)$ for brevity) and $\der_1 (a_2 {\cdot} k_1 \der_2 \calG_{21})=-\der_1 (a_2 {\cdot} P_2)$, the bosonic contractions in $\even[4]_\textup{2-bil}$ can be rewritten as a total derivative that vanishes upon integration
\begin{equation}
\left( k_1 {\cdot} a_2 \der_1 \der_2 \calG_{12} - \ap k_1{\cdot} P_1 a_2{\cdot} P_2 \right)\P_4 =
\der_1 (a_2 {\cdot} P_2) \P_4+\der_1 \P_4 a_2 {\cdot} P_2=
\der_1(a_2 {\cdot} P_2 \P_4)
\end{equation}

\subsection{Odd spin structure, four bilinears}

Four fermionic bilinears allow three types of contractions. 

First one can absorb the four zero modes at two points (for example $z_1$ and $z_2$):
\begin{equation}
\begin{split}
\odd[4]_\textup{4-bil,2}&= c_1^\textup{GSO} \sum_\textup{pairs} \braket{{:}k_1{\cdot}\y_0\,a_1 {\cdot} \y_0{:}\,{:}k_2{\cdot}\y_0\,a_2 {\cdot} \y_0{:}\,{:}k_3 {\cdot}\y\, a_3{\cdot} \y{:}\,{:}k_4 {\cdot}\y \, a_4 {\cdot} \y{:}} = \\
&=\dfrac{1}{4} \ap^4 \calC_{\calN=1} \sum_\textup{pairs} (\tilde{f}_1 f_2) (f_3 f_4) S_{34}^2 \P_4(z_i,k_i)
\end{split}
\end{equation}

Second, one can absorb two zero modes in a point and the others in two separate points. There are twelve ways to do this:
\begin{equation}
\begin{split}
\odd[4]_\textup{4-bil,1}&= c_1^\textup{GSO} \sum_\textup{cyclic} \sum_\textup{subcyclic} \sum_\textup{swaps} \braket{{:}k_1{\cdot}\y_0\,a_1 {\cdot} \y_0{:}\,{:}k_2{\cdot}\y_0\,a_2 {\cdot} \y{:}\,{:}k_3 {\cdot}\y_0\, a_3{\cdot} \y{:}\,{:}k_4 {\cdot}\y \, a_4 {\cdot} \y{:}} = \\
&=\ap^4 \calC_{\calN=1} \sum_\textup{cyclical} \sum_\textup{conn} (\tilde{f}_1 f_2 f_4 f_3) S_{24}S_{43} \P_4(z_i,k_i)
\end{split}
\end{equation}

Third, one can absorb the zero modes in four different points:
\begin{equation}
\begin{split}
\odd[4]_\textup{4-bil,0}&= c_1^\textup{GSO} \sum_\textup{swaps} \braket{{:}k_1{\cdot}\y_0\,a_1 {\cdot} \y{:}\,{:}k_2{\cdot}\y_0\,a_2 {\cdot} \y{:}\,{:}k_3 {\cdot}\y_0\, a_3{\cdot} \y{:}\,{:}k_4 {\cdot}\y_0 \, a_4 {\cdot} \y{:}} = \\
&=-2\ap^4 \calC_{\calN=1} \sum_\textup{disconn} \levi_{\m_1 \m_2 \m_3 \m_4} (f_1 f_2)^{\m_1 \m_2} (f_3 f_4)^{\m_3 \m_4} S_{12}S_{34}\Pi_4 (z_i,k_i)
\end{split}
\end{equation}

\subsection{Odd spin structure, three bilinears}
With three bilinears, one has three ways to absorb zero modes and the contractions yield
\begin{equation}
\begin{split}
\odd[4]_\textup{3-bil}= &-i c_1^\textup{GSO} \sum_\textup{cyclical} \sum_\textup{swaps} \braket{{:}a_1{\cdot}\der X {:}\,{:}k_2{\cdot}\y_0\,a_2 {\cdot} \y_0{:}\,{:}k_3 {\cdot}\y_0\, a_3{\cdot} \y{:}\,{:}k_4 {\cdot}\y_0\, a_4 {\cdot} \y{:}} = \\
&=-\ap^4 \calC_{\calN=1} \sum_\textup{cyclical} a_1{\cdot} P_1 \sum_{i=2}^4 (\tilde{f}_i f_{i+1} f_{i+2}) S_1(z_{i+1\, i+2}) \P_4(z_i,k_i)
\end{split}
\end{equation}

\subsection{Odd spin structure, two bilinears}

With two fermionic bilinears, there are six ways to absorb the four fermionic zero-modes 
\begin{equation}
\begin{split}
\odd[4]_\textup{2-bil}& =-c_1^\textup{GSO}\sum_\textup{pairs} \braket{ {:}a_1 {\cdot} \der X_1{:}\,{:}a_2 {\cdot} \der X_2{:}\,{:}k_3 {\cdot} \y_0\, a_3 {\cdot} \y_0 {:} \, {:}k_4 {\cdot} \y_0 \, a_4 {\cdot} \y_0{:}}=\\
&=-\dfrac{1}{2}\ap^3 \calC_{\calN=1} \sum_\textup{pairs} \left(a_1{\cdot} a_2 \der_1 \der_2 \calG_{12} -\ap a_1{\cdot} P_1 a_2{\cdot} P_2 \right) (\tilde{f}_3 f_4) \P_4(z_i,k_i) 
\end{split}
\end{equation}

\section{Simplifying 4-pt amplitudes}
\label{sect:simplifying_four_pt_amplitude}

Let us now simplify the above results and show that non MHV amplitudes vanish. We will also identify the regions of the integration domain that generically expose singularities and later on discuss which (tadpole) conditions the brane configurations must satisfy in order to cancel or mitigate the singular behaviours.

After analyzing the symmetry properties of the integration variables, that allow to manipulate the integrands and reduce the number of independent contributions, we will study the three independent helicity configurations and check that 
$\cA({+}{+}{+}{+}) =0 $ and $\cA({-}{+}{+}{+}) =0$.

The $\calF$-term is proportional to $F^4$ thus reproduces the MHV structure, so we will focus on the $\calE$-term.

\subsection{$\calA_4^\textup{1-loop}[1^+,2^+,3^+,4^+]=0$}
This case is the most laborious because none of the traces over the Lorentz indices of the $f_i$ vanish.

The six terms arising from contractions of two bilinears are separately gauge invariant, thus we can always choose $q_i=q_j=q$ and get $a_i{\cdot} a_j=0$ and fix $q$ so as to make some other product between momenta and polarizations vanish. For example one can compute $a_1 {\cdot} P_1 a_2{\cdot} P_2$ with the choice $q_1=q_2=k_4$ and get
\begin{equation}
a_1^+ {\cdot} P_1 a_2^+ {\cdot} P_2 (f_3^+ f_4^+)=-\dfrac{1}{2} [12]^2 (S_{12}-S_{13})(S_{21}-S_{23})(f_3^+ f_4^+)=
-\dfrac{1}{2} [12]^2 [34]^2 \left(S_{12}^2+\W_{123}\right)
\end{equation}
For brevity we define 
\begin{equation}
\label{eq:defOmega}
\W_{123}=S_{12}S_{23}+S_{23}S_{31}+S_{31}S_{12}
\end{equation}
Collecting the various Lorentz invariant structures yields
\begin{equation}
\even[4]_\textup{2-bil}=-\dfrac{1}{4}\calE_\calN \sum_\textup{disconn} (f_1^+ f_2^+)(f_3^+ f_4^+) \left(S_{12}^2+S_{34}^2+\W_{123}+\W_{134}\right)
\end{equation}
Using Schouten's identity traces with two and four $f$'s can be related
\begin{equation}
\label{eq:traces_relations}
(f_1^+ f_2^+)(f_3^+ f_4^+){=}[12]^2[34]^2{=}[12][23][34][41]{+}[12][24][43][31]{=}2(f_1^+ f_2^+ f_3^+ f_4^+){+}2(f_1^+ f_2^+ f_4^+ f_3^+)
\end{equation}
One can easily obtain two more similar formulae permuting the external legs. These formulae can be inverted to give
\begin{equation}
(f_1^+ f_2^+ f_3^+ f_4^+)=\dfrac{1}{4} \left[(f_1^+ f_2^+)(f_3^+ f_4^+)-(f_1^+ f_3^+)(f_2^+ f_4^+)+(f_1^+ f_4^+)(f_2^+ f_3^+)\right]
\end{equation}
We rewrite all the traces in terms of single traces of four $f$'s and obtain
\begin{equation}
\even[4]_\textup{2-bil}=-\dfrac{1}{2}\calE_\calN \sum_\textup{conn} (f_1^+ f_2^+ f_3^+ f_4^+) \left(S_{12}^2+S_{34}^2+S_{14}^2+S_{23}^2+2\W_{123}+2\W_{134}\right) \P_4(z_i,k_i)
\end{equation}

The three-bilinear term can be simplified using a cyclic gauge choice, for example $q_i=k_{i+2}$, and momentum conservation $a_i^+{\cdot} P_i= a_i{\cdot} k_{i+1} (S_{i\,i+1}+ S_{i+3\,i})$. With this choice all the kinematic factors become equal $a_i^+{\cdot} k_{i+1} (f_{i+1}^+ f_{i+2}^+ f_{i+3}^+)= [12][23][34][41]/2=(f_1^+ f_2^+ f_3^+ f_4^+)$. Thus one finds
\begin{equation}
\even[4]_\textup{3-bil}= -(f_1^+ f_2^+ f_3^+ f_4^+) \calE_\calN \sum_i \left(S_{i\,i+1}+ S_{i+3\,i}\right) \w_{i+1\,i+2\,i+3} \P_4(z_i,k_i)
\end{equation}
Expanding the sum we obtain
\begin{equation}
\even[4]_\textup{3-bil}= -(f_1^+ f_2^+ f_3^+ f_4^+) \calE_\calN \left[4\left(S_{12}S_{34}+S_{23}S_{41}\right)+\sum_\textup{cyclic} \left(2S_{41}S_{12}-S_{12}S_{24}-S_{24}S_{41}\right)\right] \P_4(z_i,k_i)
\end{equation}
The terms $S_{12}S_{34}+S_{23}S_{41}$ can be rewritten in four ways using partial integrations, for example choosing the tern $(412)$:
\begin{equation}
S_{12}S_{34}+S_{23}S_{41}=
-2 S_{41}S_{12}-\dfrac{u}{s}\left(S_{41}S_{12}+S_{12}S_{24}\right)-\dfrac{u}{t}\left(S_{41}S_{12}+S_{24}S_{41}\right)
\end{equation}
To obtain the other three it's enough to perform a cyclic permutation on the indices $(412)$. Replacing last equation in $\even[4]_\textup{3-bil}$ one gets
\begin{equation}
\even[4]_\textup{3-bil}{=} (f_1^+ f_2^+ f_3^+ f_4^+) \calE_\calN \sum_\textup{cyclic}\left[\left(S_{12}S_{24}{+}S_{24}S_{41}\right)\!{+}\dfrac{u}{s}\left(S_{41}S_{12}{+}S_{12}S_{24}\right)\!{+}\dfrac{u}{t}\left(S_{41}S_{12}{+}S_{24}S_{41}\right)\right] \P_4(z_i,k_i)
\end{equation}
The ratios of $s,t,u$ can be used to transform the traces of $f$'s into one another according to
\begin{equation}
u(f_1^+ f_2^+ f_3^+ f_4^+) =t(f_1^+ f_2^+ f_4^+ f_3^+) =s(f_1^+ f_3^+ f_2^+ f_4^+)
\end{equation}
that can be easily proved using the helicity formalism and momentum conservation. Thus one gets
\begin{equation}
\begin{split}
& (f_1^+ f_2^+ f_3^+ f_4^+) \left(S_{24}(S_{12}+S_{41})+\dfrac{u}{t} S_{41}(S_{12}+S_{24})+\dfrac{u}{s} S_{12}(S_{41}+S_{24})\right)= \\
& (f_1^+ f_2^+ f_3^+ f_4^+) S_{24}(S_{12}+S_{41})+(f_1^+ f_2^+ f_4^+ f_3^+) S_{41}(S_{12}+S_{24})+(f_1^+ f_3^+ f_2^+ f_4^+)S_{12}(S_{41}+S_{24})
\end{split}
\end{equation}
that can be rewritten as a ``connected" sum
\begin{equation}
\even[4]_\textup{3-bil} = \ap^4 \calE_\calN \sum_\textup{conn} (f_1^+ f_2^+ f_3^+ f_4^+) \sum_\textup{cyclic}\left(S_{12}S_{24}+S_{24}S_{41}\right) \P_4(z_i,k_i)
\end{equation}

It remains to simplify terms arising with four bilinears. We peruse (\ref{eq:traces_relations}) to rewrite disconnected terms in terms of traces of four $f$'s and use the identity $\calP_{24}+\w_{234}\w_{413}=\calP_{13}+\w_{123}\w_{341}$ to get
\begin{equation}
\even[4]_\textup{4-bil,$\calE$}=\sum_\textup{conn} (f_1^+ f_2^+ f_3^+ f_4^+) \left[\dfrac{1}{2}\left(\calP_{12}+\calP_{34}+\calP_{14}+\calP_{23}\right)+\calP_{13}+\w_{123}\w_{341}\right] \P_4(z_i,k_i)
\end{equation}
Expanding the products $\w_{123} \w_{341}$ and using partial integration one gets
\begin{equation}
\begin{split}
\w_{123} \w_{341}&{=}{-}S_{13}^2{+}S_{12}S_{34}{+}S_{23}S_{41}{-}(S_{21}S_{13}{+}S_{13}S_{32}){-}(S_{13}S_{34}{+}S_{41}S_{13}){+}S_{41}S_{12}{+}S_{23}S_{34}{=}\\
&{=}{-}S_{13}^2{-}(S_{21}S_{13}{+}S_{13}S_{32}){-}(S_{13}S_{34}{+}S_{41}S_{13}){-}\dfrac{u}{s} (S_{12}S_{24}{+}S_{41}S_{12}){-}\dfrac{u}{t} (S_{42}S_{23}{+}S_{23}S_{34})
\end{split}
\end{equation}
Recalling equation (\ref{eq:traces_relations}), the ratios $u/t$ and $u/s$ produce further mixing among the various Lorentz invariant structures:
\begin{equation}
\even[4]_\textup{4-bil,$\calE$}{=}
\sum_\textup{conn} (f_1^{+} f_2^{+} f_3^{+} f_4^{+}) \left[\dfrac{1}{2}\left(\calP_{12}{+}\calP_{34}{+}\calP_{14}{+}\calP_{23}\right){-}2 \calY_{13}{-}\sum_\textup{cyclic}\left(S_{12}S_{24}{+}S_{24}S_{41}\right) \right] \P_4(z_i,k_i)
\end{equation}
where
\begin{equation}
\calY(z_{ij})=-2\left[\calP(z_{ij})-S^2 (z_{ij})\right]
\end{equation}
Assembling all the terms and rewriting $\calP-S^2$ in terms $\calY$\footnote{See \cite{Broedel:2014vla} for more details on relations between elliptic functions and string one-loop amplitudes.} one finally gets 
\begin{equation}
\even[4] = -\calE_\calN \sum_\textup{conn}(f_1^+ f_2^+ f_3^+ f_4^+) 
\left[\calY_{12}+\calY_{34}+\calY_{14}+\calY_{23}+2\calY_{13}+\W_{123}+\W_{134} \right]\P_4(z_i,k_i)
\end{equation}
This expression vanishes using a generalized version of Fay `trisecant' identity \cite{Fay:fay}:
\begin{equation}
\label{eq:fay_identity}
\W_{123}=S_{12}S_{23}+S_{23}S_{31}+S_{31}S_{12}=-\calY_{12}-\calY_{23}-\calY_{31}
\end{equation}

Let us now verify that the amplitude vanishes in the odd spin structure too. As usual only $\calN=1$ sectors contribute. The two and three bilinears terms are the same as the sum over even spin structures, up to a by-now familiar constant:
\begin{gather}
\odd[4]_\textup{2-bil}=-\dfrac{\ap^4}{2}(i\calC_{\calN=1}) \sum_\textup{conn} (f_1^+ f_2^+ f_3^+ f_4^+) \left(S_{12}^2+S_{34}^2+S_{14}^2+S_{23}^2+2\W_{123}+2\W_{134}\right) \P_4(z_i,k_i) \\
\odd[4]_\textup{3-bil} =\ap^4 (i\calC_{\calN=1}) \sum_\textup{conn} (f_1^+ f_2^+ f_3^+ f_4^+) \sum_\textup{cyclic}\left(S_{12}S_{24}+S_{24}S_{41}\right) \P_4(z_i,k_i)
\end{gather}
Let us then consider the terms with four bilinears. The term $ \odd[4]_\textup{4-bil,2}$ is similar to the disconnected part of four bilinears in the even spin structures:
\begin{equation}
\begin{split}
\odd[4]_\textup{4-bil,2}= \dfrac{\ap^4}{2} (i\calC_{\calN=1}) \sum_\textup{conn} (f_1^+ f_2^+ f_3^+ f_4^+) \left(S_{12}^2+S_{34}^2+S_{14}^2+S_{23}^2\right) \P_4(z_i,k_i)
\end{split}
\end{equation}
The term $\odd[4]_\textup{4-bil,1}$ yields
\begin{equation}
\begin{split}
\odd[4]_\textup{4-bil,1} &= \ap^4 (i\calC_{\calN=1}) \sum_\textup{cyclical} \sum_\textup{conn} (f_1^+ f_2^+ f_4^+ f_3^+) S_{34}S_{42} \P_4(z_i,k_i)=\\
&=\ap^4 (i\calC_{\calN=1}) \sum_\textup{cyclical} \sum_\textup{conn} (f_1^+ f_2^+ f_3^+ f_4^+) S_{21}S_{14} \P_4(z_i,k_i)
\end{split}
\end{equation}
The more problematic term is $\odd[4]_\textup{4-bil,0}$ but it can be computed using the decomposition of $f$ in definite helicity parts, detailed in appendix \ref{sect:spinor_helicity_formalism},
\begin{equation}
\begin{split}
\odd[4]_\textup{4-bil,0}&=
-2 \ap^4 \calC_{\calN=1} \sum_\textup{disconn} \levi_{\m_1 \m_2 \m_3 \m_4} (f_1^+ f_2^+)^{\m_1 \m_2} (f_3^+ f_4^+)^{\m_3 \m_4} S_{12}S_{34}=\\
%&= -2 \ap^4 (i  \calC_{\calN=1}) \sum_\textup{disconn} \tensor{\levi}{^i ^j _m} \tensor{\levi}{^k ^l _n} \Tr \bar{\S}^m \bar{\S}^n (f_1^+)_i (f_2^+)_j (f_3^+)_k (f_4^+)_l S_{12}S_{34}=\\
&=-\dfrac{\ap^4}{2}(i\calC_{\calN=1}) \sum_\textup{disconn} (f_1^+ f_2^+) (f_3^+ f_4^+) \left(S_{13}S_{24}-S_{41}S_{23}\right)
\end{split}
\end{equation}
Using equation (\ref{eq:traces_relations}) the expression becomes
\begin{equation}
\odd[4]_\textup{4-bil,0}=\ap^4 (i \calC_{\calN=1}) \sum_\textup{conn} (f_1^+ f_2^+ f_3^+ f_4^+) \left(S_{12}S_{34}+S_{41}S_{23}\right)
\end{equation}
Singularly the terms $S_{12}S_{34}$ and $S_{41}S_{23}$ can be transformed in $-\W_{234}$ and $-\W_{214}$ using partial integration and mixing of Lorentz invariant but reducible structures:
\begin{equation}
\begin{split}
\sum_\textup{conn} (f_1^+ f_2^+ f_3^+ f_4^+) &S_{12}S_{34}= -\sum_\textup{conn} (f_1^+ f_2^+ f_3^+ f_4^+) \left[S_{23} S_{34}+\dfrac{u}{s}\left(S_{23}S_{34}+S_{34}S_{42}\right)\right]=\\
&\!\!\!\!\!\!\!\!\!\!\!\!=-\sum_\textup{conn} (f_1^+ f_2^+ f_3^+ f_4^+) \left[S_{23} S_{34}+\dfrac{u}{s}\left(\W_{234}-S_{42}S_{23}\right)\right]=\\
&\!\!\!\!\!\!\!\!\!\!\!\!=-\sum_\textup{conn} (f_1^+ f_2^+ f_3^+ f_4^+) \left[S_{23} S_{34}+\W_{234}-S_{43}S_{32}\right]=
-\sum_\textup{conn} (f_1^+ f_2^+ f_3^+ f_4^+) \W_{234}
\end{split}
\end{equation}
Similarly for $S_{41}S_{23}$. The final result is
\begin{equation}
\odd[4]_\textup{4-bil,0}=-\ap^4 (i \calC_{\calN=1}) \sum_\textup{conn} (f_1^+ f_2^+ f_3^+ f_4^+) \left(\W_{234}+\W_{214}\right)
\end{equation}
Summing all the terms, the contribution of the odd spin structure vanishes too. Contrary to two or three point amplitudes, contributions from even and odd spin structure are not simply proportional. Comparing even and odd spin structures $\odd[4]_\textup{2-bil}$, $\odd[4]_\textup{3-bil}$ and $\odd[4]_\textup{4-bil,2}$ are substantially equal to the even ones. $\odd[4]_\textup{4-bil,0}$ reproduces the same terms as in $\w \w$. We note that $S^2$ takes the place of Weierstrass function thus all the structures $\calP-S^2$ vanish. $\odd[4]_\textup{4-bil,1}$ has no counterpart in even sector.

\subsection{$\calA_4^\textup{1-loop}[1^-,2^+,3^+,4^+]=0$}
In this case, all the traces with $f_1^-$ are zero
\begin{equation}
(f_1^- f_i^+)=0 \quad (f_1^- f_i^+ f_j^+)=0 \quad (f_1^- f_i^+ f_j^+ f_k^+)=0
\end{equation}
As a result the $(ffff)$ terms and the four bilinears terms vanish. Only one term from three bilinears survives and three terms from two bilinears:
\begin{equation}
\even[4] {=}  {-}\ap^4 a_1^{-}{\cdot} P_1  \calE_\calN  \left[ (f_2^{+} f_3^{+} f_4^{+})\w_{234} {-}\dfrac{1}{2} a_2^{+} {\cdot}P_2 (f_3^{+} f_4^{+}) {-}\dfrac{1}{2} a_3^{+} {\cdot} P_3 (f_2^{+} f_4^{+}){-}\dfrac{1}{2} a_4^{+} {\cdot} P_4 (f_2^{+} f_3^{+}) \right] \P_4(z_i,k_i)
\end{equation}
We can compute the terms with two bilinears using the gauge choice $q_2=q_3=q_4=k_1$ and $q_1=k_3$ and obtain:
\begin{equation}
a_2^+{\cdot} P_2= a_2^+{\cdot} k_3 (S_{23}+S_{42}) \quad,\quad  a_3^+{\cdot} P_3 = a_3^+{\cdot} k_4 (S_{34}+S_{23})\quad,\quad a_4^+ {\cdot}P_4 = a_4^+ {\cdot}k_2 (S_{42}+S_{34})
\end{equation}
As $a_2^+{\cdot} k_3 (f_3^+ f_4^+)$, $a_3^+ {\cdot}k_4 (f_2^+ f_4^+)$ and $a_4^+{\cdot} k_2 (f_2^+ f_3^+)$ are all equal to $(f_2^+ f_3^+ f_4^+)$, we can factor this out and the sum vanishes
\begin{equation}
\even[4] {=}-\ap^4 \calE_\calN a_1^-{\cdot} P_1 (f_2^{+} f_3^{+} f_4^{+}) \left[ S_{23}{+}S_{34}{+}S_{42} -\dfrac{1}{2}\left(S_{23}{+}S_{42}{+}S_{34}{+}S_{23}{+}S_{42}{+}S_{34} \right) \right] \P_4(z_i,k_i){=}0
\end{equation}
In the odd spin structure $\odd[4]_\textup{2-bil}$ and $\odd[4]_\textup{3-bil}$ are substantially equal to their even counterparts, in fact by direct computation we obtain the same result with the usual replacement of $\calE_\calN$ with $i \calC_{\calN=1}$, thus the odd spin structure contribution vanishes too. So we conclude that the whole result is zero.

\subsection{$\calA_4^\textup{1-loop}[1^-,2^-,3^+,4^+]\neq 0$}
In this case, terms with traces of three $f$’s vanish. The only non-vanishing irreducible traces are $(f_1^- f_2^-)=-\<12\>^2$, $(f_3^+ f_4^+)=-[34]^2$ and
\begin{equation}
(f_1^- f_2^- f_3^+ f_4^+)=(f_1^- f_3^+ f_2^- f_4^+)=\dfrac{1}{4}\<12\>^2 [34]^2=-F^4_{--++}
\end{equation}
thanks to $f_i^+ f_j^- = f_j^- f_i^+$. For brevity we use $F^4$ instead of $F^4_{--++}$. For terms with two bilinears extra simplifications take place with the gauge choices $q_1 = q_2 = k_3,k_4$ and $q_3 = q_4 = k_1,k_2$, while terms from contractions of three bilinears vanish in this case. The $\calF$-term is non zero
\begin{equation}
\even[4]_\calF= \dfrac{1}{2} F^4 \P_4(z_i,k_i)
\end{equation}

Contributions derived from two bilinear contractions can be reduced to two terms:
\begin{equation}
\even[4]_\textup{2-bil}=\dfrac{\calE_\calN}{2} (f_3^+ f_4^+) a_1^-{\cdot} P_1 a_2^- {\cdot} P_2 \P_4(z_i,k_i) +\dfrac{\calE_\calN}{2}(f_1^- f_2^-) a_3^+{\cdot} P_3 a_4^+ {\cdot}P_4 \P_4(z_i,k_i)
\end{equation}
Using the gauge choices $q_1 = q_2 = k_4$, one finds $a_1^-{\cdot} P_1 a_2^-{\cdot} P_2=-1/2 (f_1^- f_2^-) (S_{12}^2 +\W_{123})$. If we choose $q_1 = q_2 = k_3$ we obtain $a_1^- {\cdot}P_1 a_2^- {\cdot}P_2 =-1/2(f_1^- f_2^-) (S_{12}^2 +\W_{412})$. Symmetrizing $a_1^- {\cdot}P_1 a_2^- {\cdot}P_2$ we have
\begin{equation}
a_1^- {\cdot}P_1 a_2^-{\cdot} P_2=-\dfrac{1}{4}(f_1^- f_2^-) \left(2S_{12}^2 +\W_{123}+\W_{412}\right)
\end{equation}
and a similar one for $a_3^+{\cdot} P_3 a_4^+ {\cdot}P_4$. Summing these two terms yields
\begin{equation}
\even[4]_\textup{2-bil}= \dfrac{1}{16}\calE_\calN F^4 \left(2S_{12}^2+2S_{34}^2+\W_{123}+\W_{234}+\W_{341}+\W_{412} \right)\P_4(z_i,k_i)
\end{equation}
The $\calE$-terms of four bilinear produce
\begin{equation}
\begin{split}
\even[4]_\textup{4-bil,$\calE$}&=-\dfrac{\ap^4}{8} F^4 \Big[2\calP_{12}+2\calP_{34}+ \calP_{13}+\calP_{24}+\calP_{14}+\calP_{23}+\\
& +\w_{123} \w_{341}+\w_{124} \w_{431}+\w_{132} \w_{241}+\w_{234} \w_{412}+\w_{243}\w_{312}+\w_{324}\w_{413} \Big] \P_4(z_i,k_i)
\end{split}
\end{equation}
Expanding and collecting the sums of $\w$’s 
\begin{equation}
\begin{split}
\w_{123} \w_{341}&+\w_{124} \w_{431}+\w_{132} \w_{241}+\w_{234} \w_{412}+\w_{243}\w_{312}+\w_{324}\w_{413}=\\
&=-S_{12}^2-S_{34}^2 -S_{13}^2-S_{24}^2-S_{14}^2-S_{23}^2-\W_{123}-\W_{234}-\W_{341}-\W_{412}
\end{split}
\end{equation}
Using formula (\ref{eq:fay_identity}) and summing all the contributions we have
\begin{equation}
\label{eq:even_mhv}
\even[4]= \dfrac{\ap^4}{8} F^4 \left[4\calF_\calN+ \calE_\calN \left(\calY_{12}+\calY_{34}-\calY_{13}-\calY_{24}-\calY_{14}-\calY_{23}\right) \right] \P_4(z_i,k_i)
\end{equation}

The contribution of the odd spin structure admits the same simplifications. The two bilinears contribution is similar to the even one except for a minus sign due to $\tilde{f}$:
\begin{equation}
\odd[4]_\textup{2-bil}= \dfrac{\ap^4}{16}(i \calC_{\calN=1}) F^4 \left(-2 S_{12}^2+2 S_{34}^2-\W_{123}-\W_{412}+\W_{143}+\W_{234} \right)\P_4(z_i,k_i)
\end{equation}
As in the $({+}{+}{+}{+})$ case, $\odd[4]_\textup{4-bil,2}$ cancels the $S^2$ in $\odd[4]_\textup{2-bil}$
\begin{equation}
\odd[4]_\textup{4-bil,2}= -\dfrac{\ap^4}{8}(i \calC_{\calN=1}) F^4 \left(-S_{12}^2+S_{34}^2 \right)\P_4(z_i,k_i)
\end{equation}
Terms with one bilinear yield
\begin{equation}
\begin{split}
\odd[4]_\textup{4-bil,1}&{=}\ap^4 \calC_{\calN{=}1} \!\!\!\sum_\textup{cyclical}\sum_\textup{conn} (\tilde{f}_1^- f_2^- f_4^+ f_3^+) S_{24} S_{43} \P_4(z_i,k_i){=} \ap^4 \calC_{\calN{=}1}\!\!\!\sum_\textup{cyclical} (\tilde{f}_1^- f_2^- f_3^+ f_4^+) \W_{234} \P_4(z_i,k_i){=}\\
&= -\dfrac{\ap^4}{4} (i\calC_{\calN=1}) F^4 \left(-\W_{234}-\W_{341}+\W_{412}+\W_{123}\right) \P_4(z_i,k_i)
\end{split}
\end{equation}
Terms with no bilinears yield
\begin{equation}
\begin{split}
\odd[4]_\textup{4-bil,0}&{=}
{-}2 \ap^4 \calC_{\calN} \!\!\!\!\sum_\textup{disconn} \!\!\!\levi_{\m_1 \m_2 \m_3 \m_4} \!(f_1^+ f_2^+)^{\m_1 \m_2} \!(f_3^+ f_4^+)^{\m_3 \m_4} \! S_{12}S_{34}\P_4(z_i,k_i){=}{-}2 \ap^4 \calC_{\calN} \levi_{\m_1 \m_2 \m_3 \m_4} (f_1^-)_{i^\prime} (f_2^-)_{j^\prime} (f_3^+)_i (f_4^+)_j \\
&\!\!\!\!{\times} \left[(\S^{i^\prime}\S^{j^\prime})^{\m_1 \m_2} (\bar{\S}^i \bar{\S}^j)^{\m_3 \m_4} S_{12}S_{34}{+}(\S^{i^\prime} \bar{\S}^i)^{\m_1 \m_3} (\S^{j^\prime} \bar{\S}^j)^{\m_2 \m_4} S_{13}S_{24}{+}(\S^{i^\prime} \bar{\S}^j)^{\m_1 \m_4} (\bar{\S}^i \S^{j^\prime})^{\m_3 \m_2} S_{14}S_{32}\right] \P_4(z_i,k_i)
\end{split}
\end{equation}
The matrices $(\S^{i^\prime} \bar{\S}^i)^{\m \n}$ are symmetric in $\m$, $\n$ thus contractions with the Levi-Civita tensor give zero. The remaining term vanishes too
\begin{equation}
\levi_{\m_1 \m_2 \m_3 \m_4}(\S^{i^\prime}\S^{j^\prime})^{\m_1 \m_2} (\bar{\S}^i \bar{\S}^j)^{\m_3 \m_4}=
-i \tensor{\levi}{^{i^\prime} ^{j^\prime} _{k^\prime}} \tensor{\levi}{^i ^j _k} \Tr (\S^{k^\prime} \bar{\S}^k)=0
\end{equation}
Summing all the terms in the odd spin structure and using the generalized Fay identity, we have
\begin{equation}
\odd[4]=\ap^4 (i \calC_{\calN=1}) F^4 \dfrac{5}{8}\left(\calY_{12}-\calY_{34}\right) \P_4(z_i,k_i)
\end{equation}
Finally
\begin{empheq}[box=%
\fbox]{align}
\label{eq:amplitude_4_p_1_l_mmpp}
\begin{split}
\calA_4^{\textup{1-loop}}[1^-,2^-,3^+,4^+]& = \dfrac{\ap^4 g_s^4}{8} F^4 \int_0^\infty \dfrac{dT}{T} \int d\m^{(4)} 
\Bigg[4\calF_\calN+5 i \calC_{\calN=1} \left(\calY_{12}-\calY_{34}\right)+\\
&+ \calE_\calN \left(\calY_{12}+\calY_{34}-\calY_{13}-\calY_{24}-\calY_{14}-\calY_{23}\right) \Bigg] \P_4(z_i,k_i)
\end{split}
\end{empheq}
A more symmetric result obtains for $\calA_4^{\textup{1-loop}}[1^-,2^+,3^-,4^+]$ that is invariant under exchanges of 1 with 3 and of 2 with 4.
 
\subsection{Permutation properties}

The integrands of four-point amplitude have interesting transformation properties under specific permutations of the variables $z_i$. The explicit integration measure in the planar/un-oriented case reads 
\begin{equation}
\int d\m^{(4)}_{1234}=\int_0^{\k{\cdot} iT/2} dz_1 \int_0^{z_1} dz_2 \int_0^{z_2} dz_3 \int_0^{z_3} dz_4 \d(z_4)
\end{equation}
Where $\k=1$ for the annulus and $\k=2$ for the M\"obius-strip, due to the double length of the boundary in the un-oriented case. The $T$ dependence of the domain can be removed changing variables to $z_i=\k{\cdot} iT \n_i/2$. The measure can be written in a form in which it is easy to recognize its permutation properties. To this end we can introduce the necessary step functions
\begin{equation}
\int d\m^{(4)}_{1234}=\left(\dfrac{iT \k}{2}\right)^4 \int_{\setR^4} d^4 \n \q(1-\n_1)\q(\n_1-\n_2) \dots \q(\n_3-\n_4) \q(\n_1)\dots \q(\n_4) \d(\n_4)
\end{equation}
We may use the arguments of the first four step functions as new integration variables $\a_i$ that read
\begin{equation}
\a_i=\n_i-\n_{i+1} \quad \text{for $i=1,2,3$} \quad , \quad \a_4= 1-\n_1
\end{equation}
The measure in the new variables is manifestly invariant under all permutations
\begin{equation}
\int d\m^{(4)}_{1234}=\left(\dfrac{iT \k}{2}\right)^4 \int_0^1 d\a_1 \dots \int_0^1 d\a_4 \d\left(1-\textstyle{\sum_{i=1}^4} \a_i\right)
\end{equation}

In the non-planar $3{-}1$ case, the measure is substantially equivalent to the one for planar three-point amplitudes, thus we can define variables similar to the previous case
\begin{equation}
\b_1=\n_1-\n_2 \quad,\quad \b_2=\n_2-\n_3 \quad,\quad \b_3=1-\n_3 \quad,\quad \b_4=\n_4
\end{equation}
The new measure is invariant under all the permutation of the first three variables
\begin{equation}
\int d\m^{(4)}_{123|4}=\left(\dfrac{iT}{2}\right)^4 \int_0^1 d\b_1 \int_0^1 d\b_2 \int_0^1 d\b_3 \q(1-\b_1-\b_2-\b_3) \int_0^1 d\b_4 \d(\b_4)
\end{equation}

Without changing variables and using the symmetry of the Chan-Paton factors, in the $2{-}2$ non-planar case  the measure can be rewritten as
\begin{equation}
\int d\m^{(4)}_{12|34} = \dfrac{1}{4}\left(\dfrac{iT}{2}\right)^4 \int_{[0,1]^4} d^4 \n \d(\n_4)=\dfrac{1}{16} \left(\dfrac{iT}{2}\right)^4 \int_{[0,1]^4} d^4 \n \sum_i \d(\n_i)
\end{equation}
The last identity following from the arbitrariness in the choice of the point that can be fixed at the origin. In this form it is clearly invariant under all the permutations of the $\n_i$ variables.

Symmetry properties of the measures can be used to simplify the computations. All world-sheet integrals assume the schematic form
\begin{equation}
\int d\m^{(4)}\, \calI(z_i,k_i) \P_4(z_i,k_i)
\end{equation}
The idea is to find the permutations that leave the Koba-Nielsen factor $\P_4(z_i,k_i)$ invariant  and use them to act on the function $\calI(z_i)$ in order to simplify it. If we explicitly write $\P_4(z_i,k_i)$ in the cases $4{-}0$, $3{-}1$ and $2{-}2$ we find
\begin{gather}
\P_{4{-}0}(z_i,k_i)=e^{-\ap k_1{\cdot} k_2(\calG_{12}+\calG_{34})-\ap k_1{\cdot} k_3(\calG_{13}+\calG_{24})-\ap k_1{\cdot} k_4(\calG_{14}+\calG_{23})}\\
\P_{3{-}1}(z_i,k_i)=e^{-\ap k_1{\cdot} k_2(\calG_{12}+\calG^T_{34})-\ap k_1{\cdot} k_3(\calG_{13}+\calG^T_{24})-\ap k_1{\cdot} k_4(\calG^T_{14}+\calG_{23})}\\
\P_{2{-}2}(z_i,k_i)=e^{-\ap k_1{\cdot} k_2(\calG_{12}+\calG_{34})-\ap k_1{\cdot} k_3(\calG^T_{13}+\calG^T_{24})-\ap k_1{\cdot} k_4(\calG^T_{14}+\calG^T_{23})}
\end{gather}
where $\calG^T(z_{12})=\calG(z_{12}+1/2)$. We note that $4{-}0$ and $2{-}2$ are invariant under permutations
\begin{equation}
g_u=\left(\substack {1\leftrightarrow 3 \\ 2\leftrightarrow 4}\right) \quad
g_t=\left(\substack {1\leftrightarrow 4 \\ 2\leftrightarrow 3}\right) \quad
g_s=\left(\substack {1\leftrightarrow 2 \\ 3\leftrightarrow 4}\right)
\end{equation}
The symmetry group is $\setZ_2 \times \setZ_2$. This permutations are also symmetries of the $4{-}0$ and $2{-}2$ measures. In the $2{-}2$ case that is evident. To see this in the $4{-}0$ case, we express $z_i$ in terms of the variables $\a_i$ and the Koba-Nielsen factor becomes
\begin{equation}
\P_{4{-}0}(\a_i,k_i)=e^{-\ap k_1{\cdot} k_2(\calG(\a_1)+\calG(\a_3))-\ap k_1{\cdot} k_3(\calG(\a_1+\a_2)+\calG(\a_2+\a_3))-\ap k_1{\cdot} k_4 (\calG(\a_2)+\calG(\a_4))}
\end{equation}
In terms of $\a_i$ the permutations  that leave the measure invariant are generated by
\begin{equation}
g_s: \: \a_1\leftrightarrow \a_3 \quad , \quad  g_u: \: \a_2\leftrightarrow \a_4 \quad , \quad g_t: \: \a_1\leftrightarrow \a_3,\a_2\leftrightarrow \a_4
\end{equation}
In the $3{-}1$ case there are no common permutations between the measure and $\Pi_{3{-}1}$.

Recalling the result (\ref{eq:amplitude_4_p_1_l_mmpp}) for the four-point amplitude, using the permutation property for the cases $4{-}0$ (including the un-oriented case) and $2{-}2$ we can identify some $\calY$ functions with one another:
\begin{equation}
\calY_{34}\sim \calY_{12}\quad \calY_{24}\sim \calY_{13}\quad \calY_{23}\sim \calY_{14}
\end{equation}
and find a simpler expression for (\ref{eq:amplitude_4_p_1_l_mmpp})
\begin{empheq}[box=%
\fbox]{align}
\calA_4^{\textup{1-loop}}[1^-{,}2^-{,}3^+{,}4^+]{=}\dfrac{\ap^4 g_s^4}{8} F^4 \!\!\int_0^\infty \!\dfrac{dT}{T} \!\int \!\!d\m^{(4)} \!
\Big[4\calF_\calN{+} \calE_\calN \left(\calY_{12}{+}\calY_{34}{-}\calY_{13}{-}\calY_{24}{-}\calY_{14}{-}\calY_{23}\right) \Big] {\P_4}(z_i,k_i)
\end{empheq}
The contribution of the odd spin structure vanishes. In the $3{-}1$ case no further simplification of (\ref{eq:amplitude_4_p_1_l_mmpp}) seems possible.

\subsection{Factorization}

In string theory, OPE of vertex operators  produce singularities that are related to factorization of the amplitudes on both massless and massive poles in 
intermediate channels. In sectors with $\cN =4$ susy no massless poles are expected to be exposed in two-particle channels of 4-point amplitudes, since 2- and 3-point `amplitudes' of massless states do not receive quantum corrections.
For four-point amplitudes, in sectors with $\cN =1,2$ susy,  one may expect factorization into sub-amplitudes of massless vectors connected by massless or massive propagators. However since the function $\calY$ has no poles, the four-point one-loop amplitude doesn't seem to factorize into two- and three-point one-loop sub-amplitudes of massless states but can only expose massive poles for generic values of the modular parameter $T$. Indeed, the series expansion of $\calY(z)$ produces
\begin{equation}
\calY(z)=-8 \left( \h_1+\dfrac{2\p}{T} \right) -8 \left( \dfrac{T^2}{10}(\h_2+3\h_1^2)+\h_1 T+\p^2 \right) \n^2+\calO (\n^4)
\end{equation}

In order to exposed singular behaviours associated to massless open and closed string states one has to consider the boundaries of the moduli space in $T$, capturing the UV ($T=0$) and IR ($T=\infty$) limits, we will address this issue after an interlude on `regular' branes, that give rise to super-conformal theories in the low energy limit \cite{Kachru:1998ys, Zoubos:2010kh}.

\subsection{\it Caveat}

Although our results for 4-point one-loop amplitudes look perfectly consistent in that they satisfy the expected Ward identities for gauge invariance and supersymmetry and show no `unphysical' singularities (violation of unitarity), additional subtleties may occur in the cases with reduced $\calN = 1,2$ supersymmetry\footnote{We would like to thank the anonymous referee for pointing out this issue.}. In contrast to the maximally supersymmetric $\calN = 4$ case, due to the presence of fermionic propagators $S_{ij}$ in the integrand, singularities in vanishing three-particle Mandelstam invariants such as $s_{123} = k_4^2 = 0$ may be exposed using a regulator that relaxes momentum conservation, following the pioneering paper by Minahan on one-loop beta functions in heterotic compactifications \cite{Minahan:1987ha}. Consistency of the procedure would require imposing $\textstyle{\sum}_{i<j} s_{ij}=0$. Yet finite contributions, resulting from $s_{123} /s_{123} = 1$ may appear that are completely absent in our approach. We believe 
that these finite contributions may be an artefact of the procedure that, though perfectly justified for 2- and 3-pt scattering amplitudes, that would vanish due to collinear kinetics for mass-less external states preventing any form of `scattering', is un-necessary in the 4-point case. In order to clarify this issue, one should carefully analyse further constraints on the 4-pt amplitudes such as their field-theory limit. This is beyond the scope of the present investigation. We would like to add that even in case the relevant field-theory limit of our amplitudes showed a finite discrepancy of the form $\ap/\ap = 1$ with available field theory results at one-loop, one can put the blame on higher spin states running in the loop that are obviously absent in standard field theories. In fact one could reverse the argument upside down or inside out and use loop corrections to probe string effects. Agreement with the field theory limit is only guaranteed at tree-level.

\section{Regular branes and super-conformal theories}
\label{sect:regular_branes}

So far we have not specified the open string vacuum configuration around which the vector boson scattering amplitude is computed. For illustrative purposes, we would like to focus on the simple but very interesting case of regular branes at a $\setZ_n$ orbifold singularity \cite{Kachru:1998ys, Zoubos:2010kh}. For given $n$ there might be several inequivalent choices (in fact at least two \ie (1,-1,0) and (1,1,-2)) for the action of the $\setZ_n$ on the three complex (six real) transverse coordinates $Z^I\approx \omega_n^{{{h}}_I} Z^I$, with $\omega_n = \exp(2\pi i/n)$ and ${{h}}_1+{{h}}_2+{{h}}_3 = 0$ (mod $n$). There are $n$ different kinds of fractional branes transforming according to the $n$ irreducible (one-dimensional) representations of $\setZ_n$. The low-energy dynamics is governed by a quiver field theory with $n$ nodes, corresponding to the $n$ gauge groups, and matter in bi-fundamental or adjoint representation, represented by arrows connecting the nodes. 

$N$ regular branes are collections of the same number $N$ of fractional branes of each kind. The resulting gauge group is $U(N)^n$. At low energies, \ie in the IR, the `anomalous' $U(1)$'s decouple and the dynamics is governed by a super-conformal field theory. The discrete Wilson line, representing the embedding of $\setZ_n$ in the Chan-Paton group is given by
\begin{equation}
\gamma = \oplus_{{{h}}=0}^{n-1} \omega_n^{{h}} {\bf 1}_{N\times N}
\end{equation}
so much so that 
\begin{equation}
\tr(\gamma^\ell) = 0 \quad \forall \ell\neq 0
\end{equation}
This is enough to guarantee that planar amplitudes for the states $\Phi^{(0)}$ surviving the orbifold projection be identical to the ones for $N$ D3-branes in flat space-time and vanish for the states that have been projected out \cite{Kachru:1998ys, Zoubos:2010kh}. 

\subsection{Tree level, disk}

This is obviously true at tree level where the amplitudes are given by  
\begin{equation}
\begin{split}
\cA_r^{\rm disk} & = \frac{1}{n^r} \sum_{{{h}}_i=0}^{n{-}1} \tr(\g^{{{h}}_1} \Phi_1 \g^{{{h}}_2} \Phi_2\ldots \g^{{{h}}_r}\Phi_r)=
\frac{1}{n^r} \sum_{{{h}}_i=0}^{n{-}1} \tr(\g^{{{h}}_1} \Phi_1 \g^{-{{h}}_1} \g^{{{h}}_1+ {{h}}_2}\Phi_2\ldots \g^{{{h}}_r}\Phi_r)=\\
&\!\!\!\!\!\!=\frac{1}{n^r} \sum_{{{h}}_i=0}^{n{-}1} \tr(\g^{{{h}}_1} \Phi_1 \g^{-{{h}}_1} \g^{{{h}}_1+ {{h}}_2}\Phi_2 \g^{-{{h}}_1- {{h}}_2} \g^{{{h}}_1+ {{h}}_2+{{h}}_3}\ldots \g^{{{h}}_r}\Phi_r) =\\
&\!\!\!\!\!\!=\frac{1}{n^r} \sum_{{{h}}_i=0}^{n{-}1} \tr(\g^{{{h}}_1} \Phi_1 \g^{-{{h}}_1} \g^{{{h}}_1+ {{h}}_2}\Phi_2 \g^{-{{h}}_1- {{h}}_2} \ldots \g^{\sum_i{{h}}_i}\Phi_r \g^{\sum_i{{h}}_i} \g^{- \sum_i{{h}}_i} )=
\tr(\Phi^{(0)}_1 \Phi^{(0)}_2 \ldots \Phi^{(0)}_r)
\end{split}
\end{equation}

At one-loop only $\cN=4$ sectors contribute, $\cN = 1$ and $\cN = 2$ sectors give zero, since for the latter the `empty' boundary would contribute $\tr(\gamma^\ell) = 0$. At higher loop, the $b{-}1$ `empty' boundaries would contribute 
$\prod_{i=1}^{b{-}1} \tr(\gamma^{\ell_i}) = 0$ unless $\ell_i = 0$ for all $i=1,\ldots b-1$.

Let us consider four-point amplitudes at one-loop.

\subsection{Planar amplitudes}

Let us consider first the planar $4{-}0$ case. For a given color ordering one has 
\begin{equation}
\cA_{4{-}0}^{\rm 1-loop} = \frac{1}{n} \sum_{{{h}}=0}^{n-1} \tr(\gamma^{{h}} t_1 t_2 t_3 t_4) \tr(\gamma^{{h}})\cA_{4{-}0}^{({{h}})}
\end{equation}
Vector bosons have both ends on the same kind of fractional brane, let us say the $\ell$-th. For this choice 
\begin{equation}
\cA_{4{-}0}^{\rm 1-loop} = \frac{N}{n} \tr_{\ell}(t_1 t_2 t_3 t_4) \sum_{{{h}}=0}^{n-1} \omega_n^{(\ell+1){{h}}}  \cA_{4{-}0}^{({{h}})}
\end{equation}
The situation is more involved for un-oriented and non-planar amplitudes, that are however suppressed at large $N$ \cite{Kachru:1998ys,Zoubos:2010kh}.

\subsection{Non-planar amplitudes}

Non-planar amplitudes differ in principle from the ones in the parent $\cN = 4$ theory, since the contributions of $\cN = 1$ and $\cN = 2$ sectors with ${{h}}\neq 0$ weighted by $\tr(\gamma^{\ell_i}t_1 \ldots) \tr(\gamma^{\ell^\prime_i}t^\prime_1 \ldots)$ are generically non-zero.
Let us focus on the two cases $2{-}2$ and $3{-}1$ in turn.

In the $2{-}2$ case one has 
\begin{equation}
\cA_{2{-}2}^{\rm 1-loop} = \frac{1}{n} \sum_{{{h}}=0}^{n-1} \tr(\gamma^{{h}} t_1 t_2) \tr(\gamma^{{h}} t_3 t_4)\cA_{2{-}2,{{h}}}^{\rm 1-loop}
\end{equation}
In the $3{-}1$ case one has 
\begin{equation}
\cA_{3{-}1}^{\rm 1-loop} = \frac{1}{n} \sum_{{{h}}=0}^{n-1} \tr(\gamma^{{h}} t_1 t_2 t_3) \tr(\gamma^{{h}} t_4)\cA_{3{-}1,{{h}}}^{\rm 1-loop}
\end{equation}
Factorization on massless intermediate closed string states accounts for the generalised St\"uckelberg mechanism giving mass to the anomalous $U(1)$'s for which $\tr(\gamma^{{h}} t_a) \neq 0$  \cite{Cvetic:1995rj,Ibanez:1998qp,Aldazabal:1998mr,Antoniadis:2002cs,Anastasopoulos:2003aj,Anastasopoulos:2006cz,Anastasopoulos:2006hn}.

\subsection{Un-oriented amplitudes}
The presence of $\Omega$-planes tends to generate local tadpoles that require a net number of fractional branes, whenever $n\neq 1$. In the case of $N$ D3's in flat space-time, there are 4 different kinds of $\Omega$3-planes one can add, depending on the quantised values of $B_2$ and $C_2$ \cite{Bianchi:1991eu, Bianchi:1997rf, Witten:1997bs}: $\Omega{3}^-$ leading to $SO(2N)$, $\widetilde{\Omega{3}}^-$ leading to $SO(2N+1)$, ${\Omega{3}}^+$ leading to $Sp(2N)$, $\widetilde{\Omega{3}}^+$ leading to $Sp(2N)'$ (with $\vartheta' =  \vartheta + \pi$). 
Only $\Omega{3}^-$ and ${\Omega{3}}^+$ admit a perturbative world-sheet description as the one used in the present analysis. 

For D3's at orbifold singularities one can balance the tadpole contribution of the $\Omega$-planes with the contribution of flavour branes \cite{Bianchi:2013gka}. The prototypical example is $N$ D3's at the un-oriented ${\setC}/{\setZ}^2$ singularity with 4 D7's and an $\Omega{7}^-$ \cite{Aharony:1998xz}. The low energy dynamics is governed by and $\cN =2$ SCFT with gauge group $Sp(2N)$ and 8 (half) hypermultiplets in the fundamental ${\bf 2N}$ and one hypermultiplets in the anti-symmetric skew-traceless tensor representation ${\bf N(2N{-}1)}$. The global symmetry is $SO(8)$. The spectra and vacuum configurations of un-oriented $\cN =1$ (super-conformal) quiver theories with flavour symmetries have been studied in some details in \cite{Bianchi:2013gka}. In principle one can study scattering amplitudes along the lines of the present analysis or even include the effect of closed-string fluxes leading to mass deformations of the quivers \cite{Bianchi:2014qma}. We refrain to do so here.

\section{UV and IR behaviours}
\label{sect:UV_IR}

In this section we will analyse the potential divergences of the 1-loop amplitudes. We are interested in studying such conditions as tadpole cancellation under which the amplitudes are finite. The amplitude we computed assume the schematic form
\begin{equation}
\calA_N^{\textup{1-loop}}= F^N \int_0^\infty \dfrac{dT}{T} \F(T) \int d\m^{(N)}\, f(\n_i,T) \P_4(\n_i,k_i,T)
\end{equation}
where $\Phi$ represents $\calF$, $\calE$ or $\calC$. For open strings the limits $T\to 0$ and $T\to \infty$ encode respectively the UV and the IR behaviours. Using modular transformations one can transform one-loop open-string amplitudes (direct channel) into tree-level closed-string exchange amplitudes (transverse channel) \cite{Sagnotti:1987tw, Bianchi:1988fr, Pradisi:1988xd, Bianchi:1988ux, Bianchi:1989du}.

\subsection{Bosonic propagators}
In all cases we must manipulate Koba-Nielsen factors and study their limits. To this end we need to study the limiting behaviours of the bosonic propagator on the annulus and the M\"obius strip.

\subsubsection{Direct channel}

For the annulus the propagator between two points, $z_1=\t \n_1$ and $z_2=\t \n_2+x$, is given by
\begin{equation}
\calG_\calA(\t\n_1,\t\n_2+x)=-2 \log \dfrac{\q_1 (\t \n_{12}+x)}{\q_1^\prime (0)}+2 \p   \n_{12}^2 \im \t
\end{equation}
where $x=0,1/2$ correspond to insertions on the same or different boundaries, respectively. For $x=0$, the logarithmic derivative of $\theta_1$ yields 
\begin{equation}
\dfrac{\q_1(z|\t)}{\q_1^\prime(0|\t)}= \dfrac{\sin \p \t \n}{\p} \prod_{n=1}^\infty \dfrac{(1-q^{n+\n})(1-q^{n-\n})}{(1-q^n)^2}= \dfrac{i}{\p} q^{-\n /2} \prod_{n=1}^\infty \dfrac{(1-q^{n+\n-1}) (1-q^{n-\n})}{(1-q^n)^2}
\end{equation}
where $q=e^{2\p i \t}$. We find convenient to define the functions  
\begin{equation}
h_\pm (\n,\t)=\prod_{n=1}^\infty \dfrac{1\pm q^{n+\n-1}}{1-q^n} \dfrac{1 \pm q^{n-\n}}{1-q^n}
\end{equation}
that satisfy $h_\pm (1-\n)=h_\pm (\n)$. For $x=1/2$ $h_+$ gets replaced by $h_-$. The propagators become
\begin{equation}
\calG_\calA(\t\n_1,\t\n_2)=-2  \log \left(\dfrac{i}{\p} q^{-\n_{12}(1-\n_{12})/2} h_-(\n_{12}) \right)\quad
\calG_\calA^T(\t\n_1,\t\n_2)=-2  \log \left(\dfrac{1}{\p} q^{-\n_{12}(1-\n_{12})/2} h_+(\n_{12}) \right)
\end{equation}
that are invariant under $\n \leftrightarrow 1-\nu$ in both cases. Using similar manipulations one can easily obtain the propagator on the M\"obius strip 
\begin{equation}
\calG_\calM(\t\n_1,\t\n_2)=-2  \log \left(\dfrac{i}{\p} q^{-\n_{12}(1-\n_{12})/2} h_-(\n_{12}) e^{-\frac{i\p \n_{12}^2}{2}} \right)
\end{equation}

\subsubsection{Transverse channel}

The transverse channel description results from the modular transformation $S$ for the annulus and $P=TST^2S$ for the M\"obius-strip \cite{Sagnotti:1987tw, Bianchi:1988fr, Pradisi:1988xd, Bianchi:1988ux, Bianchi:1989du}. Denoting by $\tilde{\t}$ the transformed modular parameter, one has
\begin{equation}
\tilde{z}_\calA= \frac{z}{\t_\calA} \quad , \quad \tilde{\t}_\calA=-\dfrac{1}{\t_\calA} \qquad {\rm and} \qquad
\tilde{z}_\calM= \frac{z}{2 \t_\calM-1} \quad , \quad \tilde{\t}_\calM=\dfrac{\t_\calM-1}{2\t_\calM-1}
\end{equation}
Defining $\ell=2/T$ and parametrizing $z=x+i\k T y /2$ and $\t=i\k T/2+(\k-1)/2$ ($\k= 1$ for the annulus and $\k= 2$ for the M\"obius strip), one finds
\begin{equation}
\tilde{\t}_\calA=i \ell \quad , \quad \tilde{z}_\calA=y-i x \ell \qquad {\rm and} \qquad
\tilde{\t}_\calM=\dfrac{1}{2}+\dfrac{i\ell}{4} \quad , \quad \tilde{z}_\calM=\dfrac{1}{2}(y-i x \ell)
\end{equation}
Under these transformations the propagator gets shifted by a function of $\t$. In the Koba-Nielsen factor, the shift is innocuous as it cancels thanks to momentum conservation.

On the transverse annulus, the propagator for points on the same boundary reads
\begin{equation}
\calG_\calA(\n_1,\n_2|\tilde{\t})=-2  \log \dfrac{\q_1(\n_{12}|\tilde{\t})}{\q_1^\prime(0|\tilde{\t})}=
-2  \log \left[\dfrac{\sin \p \n_{12}}{\p} g_-(\n_{12})\right]
\end{equation}
where
\begin{equation}
g_\pm(\n,\t)=\prod_{n=1}^\infty \dfrac{1\pm e^{2\p i \nu}q^n}{1-q^n} \dfrac{1\pm e^{-2\p i \nu}q^n}{1-q^n}
\end{equation}
For points on different boundaries, $\q_1$ gets replaced by $\q_2$:
\begin{equation}
\calG_\calA(\n_1,\n_2+1/2|\t)=\calG_\calA^T(\t \n_1,\t \n_2|\t)= -2  \log \dfrac{\q_2 (\n_{12})}{\q_1^\prime (0)} +2 \p   \n_{12}^2 \im \t
\end{equation}
Under $S$ modular transformations the propagator becomes
\begin{equation}
\calG_\calA^T(\n_1,\n_2|\tilde{\t})=-2  \log \dfrac{\q_2(\n_{12}|\tilde{\t})}{\q_1^\prime(0|\tilde{\t})}=
-2  \log \left[\dfrac{\cos \p \n_{12}}{\p} g_+(\n_{12})\right]
\end{equation}
For the M\"obius strip the propagator is
\begin{equation}
\calG_\calM(\n_1,\n_2|\tilde{\t}_\calM)=-2  \log \left[ \dfrac{\sin \p \n_{12}/2}{\p} g_+\left(\frac{\n_{12}}{2},\tilde{\t}_\calM\right) \right]
\end{equation}

\subsection{Functions $\calF_\calN$, $\calE_\calN$, $\calC_\calN$ and $\calY$}

In addition to the propagators, one needs the limiting behaviours of the functions $\calF_\calN$, $\calE_\calN$, $\calC_\calN$ and $\calY$. For simplicity we will only consider configurations of branes at orbifold singularities, in particular `regular' branes \cite{Kachru:1998ys, Zoubos:2010kh}, thus we can set $\e^I_{ab}=0$ in $u^I_{ab}$.

\subsubsection{$\calN=4$ sectors}
It is easy to see that $\calE_{\calN=4}=0$ due to Riemann identity. For the same reason $\calF_{\calN=4}$ is simply proportional to $\L^{(6)}/T^2$ with
\begin{equation}
\L^{(6)}=\sum_{\{p\}} e^{- 2 \p  \im \t \ap p^2/R^2}
\end{equation}
for D9-branes. Using Poisson resummation, for a lattice with dimension $2r$ one finds
\begin{equation}
\L^{(2r)}(\t_2)=\left(\dfrac{2\ap \t_2}{R^2} \right)^{r} \L^{(2r)}(\t_2^{-1})
\end{equation}
Using T-duality along all 6 internal directions one gets D3-branes and the lattice sum becomes
\begin{equation}
\L^{(6)}_\textup{D3}=\sum_{\{w\}} e^{- 2 \p \im \t w^2 \tilde{R}^2/\ap }
\end{equation}
Introducing a non zero separation $\D x$ between the D3-branes one can regularize IR divergences. $\D x$ is related to the mass of the lowest states by $\D x= \ap  M$. The separation $\D x $ can be chosen in many ways in the $\calN=4$ sector: the branes are parallel in all the six compact dimension thus they can be displaced along one, two or three complex dimensions ${{{{D_s}}}}$.
\begin{equation}
\L^{(6)}_\textup{D3}=\sum_{\{w\}} e^{- 2 \p  \im \t (w \tilde{R}+\D x)^2/\ap } \qquad, \qquad \L^{(6)}_\textup{D9}=\sum_{\{p\}} e^{- 2 \p  \im \t \ap (p+a)^2/R^2} 
\end{equation}
where the second formula is for the D9-branes description with $a$ the Wilson line related to branes' separation by $\D x= a \ap/R$. In the limit $T\to \infty$ the behaviour of $\L^{(6)}$ is dominated by the exponential
\begin{equation}
\L^{(6)}_\textup{D9}\xrightarrow{T\to \infty} {{b_p}} e^{-2\p \ap \im \t \min_p (p+a)^2/R^2 }
\end{equation}
where ${{b_p}}$ accounts for possible degeneracies. For $|a|<1/2$ the minimum corresponds to $p=0$ and its value is $M^2$. In the transverse channel the limit $\ell \to \infty$ produces
\begin{equation}
\L^{(6)}_\textup{D9}=\sum_{\{p\}} e^{- 2 \p \ap (p+a)^2/\ell R^2} \xrightarrow{l\to \infty} \left(\dfrac{\ell R^2}{2\ap}\right)^3
\end{equation}
In the partial decompactification limit $R \to \infty$ with separation along ${{D_s}}$ directions, one has
\begin{equation}
\L^{(6)}=\dfrac{V_6}{(2 \p \ap \im \t)^{3-{{D_s}}}} e^{-2\p \ap \im \t M^2 }
\end{equation}
where $V_6$ is the regulated volume of $\setR^6$. The limits in this case yield
\begin{equation}
\L^{(6)}_\textup{D9}(R=\infty) \xrightarrow{T\to \infty} \dfrac{V_6}{(\p \ap T)^{3-{{D_s}}}} e^{-\p \ap T M^2 } \qquad, \qquad
\L^{(6)}_\textup{D9}(R=\infty) \xrightarrow{\ell\to \infty} V_6\left( \dfrac{\ell}{2\p \ap}\right)^{3-{{D_s}}}
\end{equation}
In conclusion the limits of $\calF_{\calN=4}$ produce
\begin{equation}
\calF_{\calN=4}\xrightarrow{T\rightarrow \infty} \dfrac{1}{4n \ap^2} \dfrac{(2\p)^4}{T^2} e^{-\p \ap T M^2} \qquad, \qquad
\calF_{\calN=4}\xrightarrow{\ell\rightarrow \infty} (2\p)^4 \dfrac{1}{4n \ap^2} \dfrac{\ell^2}{\k^4} \left(\dfrac{\ell R^2}{2\ap} \right)^3
\end{equation}

\subsubsection{$\calN=2$ sectors}
In this case $\calE_{\calN=2}$ is non-zero, in fact 
\begin{equation}
\calF_{\calN=2}=-\calE_{\calN=2} \calP(u) \qquad \qquad \calE_{\calN=2}=(2\p)^2 \calX_{\calN=2}=\dfrac{ (2\p)^2 \L^{(2)}(\t) I^{(4)}}{4 n (2\ap \im \t)^2}
\end{equation}
$I^{(4)}$ is a constant in the limits $\ell, T \to \infty$, $\L^{(2)}$ is substantially equal to $\L^{(6)}$ with the restriction that brane separation can take place only along one complex direction.
\begin{equation}
\L^{(2)}_\textup{D9}\xrightarrow{T\to \infty} e^{-2\p \ap \im \t M^2} \quad
\L^{(2)}_\textup{D9}\xrightarrow{l\to \infty} \dfrac{\ell R^2}{2\ap}
\end{equation}
In the amplitudes terms like $\calE_{\calN=2}$, often appear in combination with Weierstrass $\calP$ function or the function $\calY$. 

Let us focus first on the $\calP(z,\t)$ and consider $z=x+i y \t_2$ ($\t_2=\im \t\propto T$) with $y=0$ and $y\neq 0$. The variable $z$ can either play the role of $u^I_{ab}={{h}} v^I_{ab} $ or of a world-sheet coordinate. In general $z=i\t_2 \n + (1/2)$. Anyway the in front of $i\t_2$ is a real number between zero and one. Expanding Weierstrass function as a power series in $q$ yields
\begin{equation}
\calP(z,\t)= \dfrac{\p^2}{\sin^2 \p z} - \dfrac{\p^2}{3} +8\p^2 \sum_{n=1}^\infty \sum_{d_n|n} q^n d_n \left[1- \cos (2\p d_n z) \right]
\end{equation}
where $d_n$ are the divisors of $n$. In the limit $T \to \infty$ the cosines diverge as $e^{2\p \t_2 d_n y}$, but this contribution is suppressed by $q^n\sim e^{-2\p \t_2 n}$ since $d_n y < n$, 
\begin{equation}
\calP(z,\t) \xrightarrow{T\to \infty} \dfrac{\p^2}{\sin^2 \p z} - \dfrac{\p^2}{3}\to \begin{cases}
-\p^2/3 & \text{if $y\neq 0$} \\
\p^2/\sin^2 (\p x) -\p^2/3 & \text{if $y=0$}
\end{cases}
\end{equation}
In the limit $\ell \to \infty$ Weierstrass function becomes
\begin{equation}
\calP(z,\t)=\tilde{\t}^2 \calP(\tilde{z},\tilde{\t}) \xrightarrow {\ell\to \infty} 
\tilde{\t}^2 \left(\dfrac{\p^2}{\sin^2 \p \tilde{z}} - \dfrac{\p^2}{3}\right) \to -\dfrac{\ell^2 \p^2}{\k^4} 
\begin{cases}
-1/3 & \text{if $x\neq 0$} \\
1/\sin^2 (\p y/\k)-1/3 & \text{if $x=0$}
\end{cases}
\end{equation}

Let us now consider the function $\calY$ only in the case $y>0$ and $x=0,1/2$. We need $S(z)$ and the series expansion of $\der_z \log \q_1$:
\begin{equation}
\der_z \log \q_1(z|\t)= \dfrac{\q_1^\prime(z|\t)}{\q_1(z|\t)}= \p\cot (\p z)+4\p \sum_{n=1}^\infty \sum_{d_n} q^n \sin (2\p d_n z)
\end{equation}
The limits $T \to \infty$ and $\ell\to \infty$ produce
\begin{equation}
S(z,\t)\xrightarrow{T\to \infty}
\begin{cases}
-i\p y \left(1/|y|{+}2 \right) & \text{if $y{\neq} 0$} \\
1/x & \text{if $y{=}0,x{\to}0$}
\end{cases}
\quad 
S(z,\t)\xrightarrow{\ell\to \infty} \dfrac{i \p \ell}{\k^2}
\begin{cases}
2i & \text{if $x{=}1/2$} \\
\cot \p y /\k & \text{if $x{=}0$}
\end{cases}
\end{equation}
Now it's easy to compute the limits for the functions $\calE_\calN$, $\calF_\calN$ and $\calY(z)$ that read
\begin{gather}
\calE_{\calN=2} \xrightarrow{T\to \infty} \dfrac{  I^{(4)} }{4n \ap^2} \dfrac{4 \p^2 }{T^2} e^{-\p \ap T M^2}\quad, \quad
\calE_{\calN=2} \xrightarrow{\ell\to \infty} \dfrac{  I^{(4)} }{4n \ap^2} \p^2 \ell^2 \dfrac{\ell R^2 }{2 \ap} \\
\calF_{\calN=2} \xrightarrow{T\to \infty} \dfrac{  I^{(4)} }{4n \ap^2} \dfrac{4 \p^4 }{T^2} \left(\dfrac{1}{\sin^2 ({{h}} \p v)}-\dfrac{1}{3}\right) e^{-\p \ap T M^2} \quad, \quad
\calF_{\calN=2} \xrightarrow{\ell\to \infty}- \dfrac{\ell^2 \p^2}{\k^4} \dfrac{  I^{(4)} }{4n \ap^2} \dfrac{\p^2 \ell^2}{3} \dfrac{\ell R^2 }{2 \ap} \\
\calY(z) \xrightarrow{T\to \infty} -4 \p^2\left(\dfrac{1}{3}+2|y|(1+|y|)\right) \quad, \quad
\calY(z) \xrightarrow{\ell\to \infty} \dfrac{\ell^2 \p^2}{\k^4}
\begin{cases}
11 & \text{if $x=1/2$}\\
2 & \text{if $x=0$}
\end{cases}
\end{gather}

\subsubsection{$\calN=1$ sectors}
This is the most laborious case, one needs to study the function
\begin{equation}
\calH(z)=\prod_{I=1}^3 \q_1(z+u^I_{ab})
\end{equation}
since $\calF_{\calN=1}$ and $\calE_{\calN=1}$ are given by
\begin{equation}
\calF_{\calN=1}=\calE_{\calN=1} \left(\dfrac{1}{6} \dfrac{\calH^{\prime \prime \prime}(0)}{\calH^\prime(0)}+3\h_1\right) \qquad
\calE_{\calN=1}=2\p \dfrac{\calH^\prime(0)}{\calH(0)} \calX_{\calN=1}
\end{equation}
Let us start with $\calE_{\calN=1}$, the ratio $\calH^\prime/\calH$ can be seen as a sum of logarithmic derivatives 
\begin{equation}
\dfrac{\calH^\prime(0)}{\calH(0)}= \textstyle{\sum_{I=1}^3}\der_z \log \q_1\big|_{z=u_I}
\end{equation}
It is easy to see that this function is a modular form of weight one. In the limits the $q$-series vanish thus only cotangents remain in the direct channel
\begin{equation}
\dfrac{\calH^\prime(0)}{\calH(0)}\xrightarrow{T \to \infty} \p \textstyle{\sum_I} \cot(\p v_I {{h}})
\end{equation}
In the transverse channel one has
\begin{equation}
\dfrac{\calH^\prime(0)}{\calH(0)}\xrightarrow{\ell \to \infty} \dfrac{i \p \ell}{\k^2} \textstyle{\sum_I} \cot (\p u_I) \to -\dfrac{\p \ell}{\k^2}\sum_I \frac{u_I}{|u_I|}
\end{equation}

$\calF_{\calN=1}$ is even more laborious to analyse
\begin{equation}
\calF_{\calN=1}=2\p \left(\dfrac{1}{6} \dfrac{\calH^{\prime \prime \prime}(0)}{\calH(0)}+3\h_1 \dfrac{\calH^\prime(0)}{\calH(0)} \right) \calX_{\calN=1}= 2\p \F(u_I) \calX_{\calN=1}
\end{equation}
One can expand the expression in brackets using the logarithm derivatives of $\q_1$, that we call $\f(z|\t)=\der_z \log \q_1(z|\t)$ for brevity:
\begin{equation}
\begin{split}
\F(u_I)&=
\prod_I \f(u_I) +\dfrac{1}{2} \sum_{I_1\neq I_2} \f(u_{I_1}) \left(\f^\prime(u_{I_2})+\f^2(u_{I_2})\right)+\\
+& \dfrac{1}{6} \sum_I \left(\f^{\prime\prime}(u_I)+3\f^\prime(u_I) \f(u_I)+\f^3(u_I)\right)+3\h_1 \sum_I \f(u_I)
\end{split}
\end{equation}
Under modular transformations the functions $\f(z)$ and $\h_1$ are modular forms of weight one and two, respectively. The function $\F(z)$ is a modular form of weight three. The limit $T\to \infty$ of $\F(u_I)$ can be computed similarly to $\calH^\prime/\calH$ and the limit $\ell\to \infty$ in the transverse channel can be taken using $\F (u)= \tilde{\t}^3 \F(\tilde{u})$
\begin{equation}
\F(z)\xrightarrow{T\to \infty} \p^3 \left(\prod_I \cot({{h}}\p v_I) -\dfrac{2}{3} \sum_I \cot ({{h}} \p v_I )\right) \quad , \quad 
\F(z)\xrightarrow{l\to \infty} -i\left( \dfrac{i\p\ell}{\k^2} \right)^3 \sum_I \frac{u_I}{|u_I|}
\end{equation}

In the $\calN=1$ case the function $\calC_\calN$ also appears that has a simple form. For future use, we list here all the limits \begin{gather}
\calE_{\calN=1}\xrightarrow{T \to \infty} \dfrac{  I^{(6)}}{4n \ap^2} \dfrac{2\p^2}{T^2} \sum_I \cot ({{h}}\p v_I) \quad, \quad
\calE_{\calN=1}\xrightarrow{\ell \to \infty} - \dfrac{  I^{(6)}}{4n \ap^2} \dfrac{ \p^2 }{2} \dfrac{\ell^3}{\k^2}  \sum_I \frac{u_I}{|u_I|}\\
\calF_{\calN=1}\xrightarrow{T \to \infty} \dfrac{  I^{(6)}}{4n \ap^2} \dfrac{2\p^4}{T^2} \textstyle{\left[\prod_I \cot({{h}}\p v_I) {-}\frac{2}{3} \sum_I \cot ({{h}} \p v_I )\right]} \,\, , \,\,
\calF_{\calN=1}\xrightarrow{\ell \to \infty} -\dfrac{  I^{(6)}}{4n \ap^2} \dfrac{\p^4 \ell^5}{2\k^6} \textstyle{{\sum_I} \frac{u_I}{|u_I|}}\\
\label{eq:C_limits}
\calC_\calN(T)= -\dfrac{  c_1^\textup{GSO} I^{(6)}}{4 n \ap^2} \dfrac{2\p^2}{T^4} \quad, \quad
\calC_\calN(\ell)= - \dfrac{  c_1^\textup{GSO} I^{(6)}}{4 n \ap^2} \dfrac{\ell^4\p^2}{8}
\end{gather}

\subsection{Two- and three-point `amplitudes'}
We now have all the tools to compute the UV and IR limits of two- and three-point `amplitudes'. 

In $\calN=4$ sectors both amplitudes are zero because $\calE_{\calN=4}=0$. This is an expected result in fact in $D=4$ SYM $\calN=4$ theory has a vanish $\beta$ function and no threshold corrections to the gauge kinetic function. 

Let us start with the two-point `amplitude' that reads
\begin{equation}
\calA_2^\textup{1-loop}[1^+,2^+]= \dfrac{g_s^2\ap^2}{2} [12]^2 \int_0^\infty \dfrac{dT}{T} \left(\calE_\calN + i \calC_{\calN=1} \right) \int d\m^{(2)}\, e^{-\ap k_1{\cdot} k_2 \calG_{12}}
\end{equation}
Despite analytic continuation of momenta, $k_1 {\cdot} k_2=0$ thus the integral over the world-sheet insertions gives $(iT/2)^2$. For this reason the color-ordered amplitude is the same in the planar and non-planar cases, they differ in the Chan-Paton factor. In the IR limit in $\calN=2$ sectors one has a logarithmic divergence due to $\L^{(2)}$
\begin{equation}
\calA_2^\textup{1-loop}[1^+,2^+]\xrightarrow{T\to \infty} -g_s^2 \dfrac{\p^2 R^2 I^{(2)}}{8 \ap n} [12]^2 \log L
\end{equation}
One can regularize IR divergences introducing brane separation in the amplitude 
\begin{equation}
\calA_2^\textup{1-loop}[1^+,2^+]\xrightarrow{T\to \infty} -g_s^2 \dfrac{\p^2 I^{(2)}}{8 n} [12]^2 \int^\infty \dfrac{dT}{T} e^{-\p \ap T M^2}
\end{equation}
In the $\calN=1$ case there are no directions to separate the branes and the integral over $T$ diverges logarithmically
\begin{equation}
\calA_2^\textup{1-loop}[1^+,2^+]\xrightarrow{T\to \infty} g_s^2\dfrac{\p^2 I^{(6)} }{n} [12]^2 
\textstyle{\sum_I} \cot({{h}} \p v_I) \log L
\end{equation}
where $L$ is the IR cut-off. We note that the odd spin structure does not contribute to the limit.

In the transverse channel we treat separately planar and un-oriented amplitudes, related to massless tadpoles and their cancellation, and the non-planar amplitude, related to the masses of anomalous $U(1)$ vector bosons. For momentum conservation $k_1{\cdot} k_2=0$ and the integral on the world-sheet in the direct channel produces $-T^2/4$ in any case, while in the transverse channel it gives $-1/\ell^2$ in the planar and non-planar case and $-16/ \ell^2$ in the un-oriented case. Summing planar and un-oriented contributions produces
\begin{equation}
\calA^\textup{1-loop}_{2{-}0}(1^+,2^+)= -\dfrac{g_s^2 \ap^2}{2} [12]^2 \left[\tr(\g^{{h}}) \tr(t_1 t_2 \g^{-{{h}}})+32 \tr(t_1 t_2 W^\W_{2{{h}}}) \right] \int_0^\infty \dfrac{d\ell}{\ell^3} \left(\calE_\calN + i \calC_{\calN=1} \right)
\end{equation}
where in the integral over $\ell$, the measure descends from $\t=\t_\calA=i\ell$. In the non-planar case
\begin{equation}
\calA^\textup{1-loop}_{1{-}1}(1^+,2^+)= -\dfrac{g_s^2 \ap^2}{2} [12]^2 \tr(\g^{{h}} t_1) \tr(t_2 \g^{-{{h}}}) \int_0^\infty \dfrac{d\ell}{\ell^3} \left(\calE_\calN + i \calC_{\calN=1} \right)
\end{equation}
The behaviour of both amplitudes is coded in color-ordered amplitudes, their limit in the direct channel for $\calN=2$ sectors reads
\begin{equation}
\calA^\textup{1-loop}_{1{-}1}[1^+,2^+] \xrightarrow{\ell \to \infty} - g_s^2[12]^2 \dfrac{  I^{(4)}}{4 n} \dfrac{\p^2}{2} \dfrac{R^2}{2 \ap} L
\end{equation}
where $L$ is the IR cutoff. In $\calN=1$ sectors the limit is dominated by the contribution of the odd spin structure
\begin{equation}
\calA^\textup{1-loop}_{1{-}1}[1^+,2^+] \xrightarrow{\ell \to \infty} i g_s^2 [12]^2 \dfrac{ c_1^\textup{GSO}   I^{(6)}}{4 n} \dfrac{L^2 \p^2}{32}
\end{equation}

Let us now consider the (color-ordered) three-point amplitude with helicities $({-}{+}{+})$ that reads
\begin{equation}
\calA_3^\textup{1-loop}[1^-,2^+,3^+]= -\dfrac{i}{\sqrt{2}}g_s \ap^2 \dfrac{[23]^3}{[12][31]} \int_0^\infty \dfrac{dT}{T} \left(\calE_\calN-i\calC_{\calN=1} \right) \ap k_2{\cdot} k_3  \int d\m^{(3)} S_{23} \P_3(z_i,k_i)
\end{equation}
As for two-point amplitudes $\calN=4$ sectors do not contribute. The form of the color-ordered amplitude is different in the planar and non-planar cases. Actually there are two non-planar cases: one with $2^+$ and $3^+$ in the second boundary, that we call $1{-}2$, and one with $1^-$ and $2^+$ in the first boundary, that we call $2{-}1$. 

Let us focus on the world-sheet integral, that we need to compute for $k_i{\cdot}k_j \to 0$. In the planar case, using the variables $\a_1=\n_{12}$, $\a_2=\n_{23}$ and $\a_3=1-\n_3$ one gets
\begin{equation}
\int \!d \m^{(3)}_{123} \Pi_3(z_i, k_i)=\t^3\!\! \int \! d\a_1 d\a_2 d\a_3\, \d(\textstyle \sum_i \a_i-1) \ap k_2{\cdot}k_3 S(\a_2 \t) e^{-\ap k_1{\cdot}k_2 \calG(\a_1 \t)-\ap k_2{\cdot}k_3 \calG(\a_2 \t) -\ap k_1{\cdot}k_3\calG(\a_3\t)}
\end{equation}
In this form it is not clear what happens when $\a_2 \to 0$ and $k_2{\cdot} k_3 \to 0$. As in \cite{Minahan:1987ha}, if we call $\e=k_2{\cdot} k_3 $, when $\e \to 0$ the leading term is finite
\begin{equation}
\lim_{\e \to 0} \left(\dfrac{iT}{2}\right)^2 \int d\a_1 d\a_3\, \d(1-\a_1-\a_3) e^{-\ap k_1{\cdot}k_2 \calG(\a_1 \t)-\ap k_1{\cdot}k_3\calG(\a_3\t)} \int d\a_2 \e (\a_2)^{\e-1} 
\end{equation}
As a result the integral gives $(iT/2)^2$. This holds true for $\calA^\textup{1-loop}_{1{-}2}[1^-,2^+,3^+]$ too, in fact using the parametrization $\b_1=\n_1$, $\b_2=\n_{23}$ and $\b_3=1-\n_3$ yields a similar integral. In the case $\calA^\textup{1-loop}_{2{-}1}[1^-,2^+,3^+]$, the amplitude does not exhibit the kinematical pole. This is in line with the Chan-Paton factor that does not involve the structure constants $f_{abc}$ but rather a product of a $\delta$ for 1 and 2 combined with an anomalous $U(1)$ factor for 3. The kinematical factor in the numerator does not cancel and makes this IR contribution vanish.

In the direct and transverse channel, the behaviour of the amplitude is very similar to the two-point amplitude in fact $\calE_\calN$ appears in the both cases. The world-sheet integral give us the same contribution in terms of modular parameter $T$ (or $\ell$). This should be interpreted in terms of the running of the field-dependent gauge couplings \cite{Anastasopoulos:2006hn} and is to be expected as a result of supersymmetry Ward identities that only allow the super-invariant $W^2$ for 2- and 3-point amplitudes, barring anomalous $U(1)$'s.

\subsection{Four-point amplitude, $4{-}0$}

Let us now study the limiting behaviours of 4-point amplitudes. In the direct channel, relying on $u=-s-t$, the Koba-Nielsen factor can be rewritten in terms of the variables $\a_i$ as
\begin{equation}
\P_{4{-}0}{=}{\exp} \!\left[\frac{\ap s}{2}\left[\calG(\a_1){+}\calG(\a_3){-}\calG(\a_1{+}\a_2){-}\calG(\a_2{+}\a_3)\right]\!{+}\frac{\ap t}{2}\left[\calG(\a_2){+}\calG(\a_4){-}\calG(\a_1{+}\a_2){-}\calG(\a_2{+}\a_3)\right]\right]
\end{equation}
Inserting in this expression the explicit form of the propagator found in the previous section, the $\a$-independent terms cancel and using $\a_1+\a_2+\a_3+\a_4=1$ many terms simplify so that
\begin{equation}
\begin{split}
\P_{4{-}0}(\a_i,k_i)=\left[\dfrac{h_- (\a_1)h_- (\a_3)}{h_- (\a_1+\a_2)h_- (\a_2+\a_3)} \right]^{-\ap s} \left[\dfrac{h_- (\a_2)h_- (\a_4)}{h_- (\a_1+\a_2)h_- (\a_2+\a_3)} \right]^{-\ap t} q^{-\ap (s \a_2 \a_4+t \a_1 \a_3)}
\end{split}
\end{equation}
In the limit $T \to \infty$, $q \to 0$ all the functions $h(\a) \to 1$ while the $s,t$ dependent $q$-exponential, using the saddle point method, behaves as
\begin{equation}
\int d^4 \a \d(1-\textstyle{\sum} \a) e^{\p T \ap (s \a_2 \a_4+t \a_1 \a_3)} \xrightarrow{T \to +\infty} e^{\frac{\p T \ap st}{4(s+t)}} 
\end{equation}
In the physical region $s>0$, $t<0$ (in the Regge limit $|s|\gg |t|$), with this condition the exponent is $-\p T \ap |t|/4$ thus the integral on $T$ is always convergent in the limit $T\to \infty$.

As in \cite{Dudas:1999gz}, if we expand the functions $h_-$ in powers of $q$. The form of the exponential suggests that one can interpret the modulus $T$ as a Schwinger parameter and the integral over the variables $\a_i$  as a Feynman parametrization of a box integral. The string amplitude can be seen as an infinite sum of massive box amplitudes.

In the transverse channel we find the usual tadpole, in fact we can obtain the same world-sheet integral rescaling the world-sheet coordinate on the M\"obius strip by a factor of two and obtaining a $2^4$ factor from $\calF_\calN$ or $\calY \calE_\calN$. In the limit $\ell \to \infty$ the Koba-Nielsen factor reduces to products or ratios of sines
\begin{equation}
\begin{split}
\calA^\textup{1-loop}_{4{-}0}&[1^-,2^-,3^+,4^+]\xrightarrow{\ell \to \infty} \dfrac{g_s^4 \ap^4}{2} F^4 \int_0^\infty \dfrac{d\ell}{\ell} \left(\calF_\calN- \dfrac{\p^2 \ell^2}{\k^4} \calE_\calN \right) \\
&\int d\m^{(4)}_{1234} \left[\dfrac{\sin \p \a_1 \sin \p \a_3}{\sin \p (\a_1+\a_2) \sin \p (\a_2+\a_3)} \right]^{-\ap s} \left[\dfrac{\sin \p \a_2 \sin \p \a_4}{\sin \p (\a_1+\a_2) \sin \p (\a_2+\a_3)} \right]^{-\ap t}
\end{split}
\end{equation}
In the $\ell \to \infty$ limit, it is useful to define a combination of $\calF$ and $\calE$ and evaluate its limit in the various sectors $\calN=1,2,4$
\begin{equation}
\label{eq:F_E_combination}
\left(2\calF_\calN+a\dfrac{\p^2 \ell^2}{\k^4} \calE_\calN \right) \xrightarrow{\ell \to \infty} 
\dfrac{\p^4 }{4n\ap^2} \dfrac{\ell^2}{\k^4}
\begin{cases}
16 (\ell R^2/2\ap)^3 & \text{if $\calN=4$} \\
(a-2/3) I^{(4)} \ell^2 (\ell R^2/2\ap) & \text{if $\calN=2$} \\
-(2+a)\dfrac{\ell^3 I^{(6)}}{2 \k^2}  \sum_I \frac{u_I}{|u_I|}& \text{if $\calN=1$}
\end{cases}
\end{equation}
In the $4{-}0$ case $a=-2$, thus the contribution to the amplitude from $\calN=1$ sectors vanishes in the limit.

The world-sheet integral can be rewritten as a derivative of the Veneziano-like amplitude wrt the string tension as in \cite{Green:2012pqa} \begin{equation}
\int d\m^{(4)}_{1234} \left[\dfrac{s_1 s_3}{s_{1+2} s_{2+3}} \right]^{-\ap s} \left[\dfrac{s_2 s_4}{s_{1+2} s_{2+3}} \right]^{-\ap t}\!\!\!\!{=}{-}\dfrac{1}{2\p^2} \int_0^1\!\! \left( \dfrac{\log x}{1-x}+\dfrac{\log(1-x)}{x}\right) x^{-\ap s} (1{-}x)^{-\ap t}
\end{equation}
As shown in \cite{Green:2012pqa}, the one-loop dilaton tadpole is proportional to the logarithmic derivative of the tree level amplitude with respect to the (inverse) tension $\ap$. 

\subsection{Four-point amplitude, $2{-}2$}
In this case one has two independent color-ordered amplitudes, in fact fixing legs 1 and 2 on a boundary and legs 3 and 4 on the other boundary, one has $\calA^\textup{1-loop}_{2{-}2}[1^-,2^-,3^+,4^+]$ and $\calA^\textup{1-loop}_{2{-}2}[1^+,2^-,3^-,4^+]$.
In the direct channel we have the same limits, $\calN=2,4$ is regularized by brane separation and $\calN=1$ by momentum flow.

In the transverse channel, one has a behaviour similar to $4{-}0$, the unique changes affect the value of the parameter $a$ that we have defined in the equation (\ref{eq:F_E_combination}): for $\calA^\textup{1-loop}_{2{-}2}[1^-,2^-,3^+,4^+]$ one has $a=-20$ while for $\calA^\textup{1-loop}_{2{-}2}[1^+,2^-,3^-,4^+]$ $a=-2$ again, thus the last amplitude vanishes for $\calN=1$. For instance, for $[1^-,2^-,3^+,4^+]$ one finds
\begin{equation}
\begin{split}
\calA^\textup{1-loop}_{2{-}2}&[1^-,2^-,3^+,4^+]\xrightarrow{\ell \to \infty} \dfrac{g_s^4 \ap^4}{4} F^4 \int_0^\infty \dfrac{d\ell}{\ell} \left(2\calF_\calN+a \dfrac{\p^2 \ell^2}{\k^4} \calE_\calN \right) \\
&\int d\m^{(4)}_{123|4} \left[\dfrac{\sin \p \b_1 \sin \p \b_3}{\cos \p (\b_1+\b_2) \cos \p (\b_2+\b_3)} \right]^{-\ap s} \left[\dfrac{\cos \p \b_2 \cos \p (\b_1+\b_2+\b_4)}{\cos \p (\b_1+\b_2) \cos \p (\b_2+\b_3)} \right]^{-\ap t}
\end{split}
\end{equation}
where $\b_1=\n_1-\n_2$, $\b_2=1-\n_1$ and similarly for $\b_3$ and $\b_4$. For $[1^+,2^-,3^-,4^+]$ one can easily adapt the formula.

\subsection{Four-point amplitude, $3{-}1$}
The non-planar amplitude $3{-}1$ has a unique independent color-ordered amplitude, $\calA^\textup{1-loop}_{3{-}1}$ $[1^-,2^-,3^+,4^+]$ with the leg 4 on the second boundary.

In the transverse channel, the contribution $\calC_\calN$ of the odd spin structure dominates in limit $\ell \to \infty$ and one gets
\begin{equation}
\begin{split}
\calA^\textup{1-loop}_{3{-}1}&[1^-,2^-,3^+,4^+]\xrightarrow{\ell \to \infty} \dfrac{g_s^4 \ap^2}{8} \p^4 \dfrac{c_1^\textup{GSO}I^{(6)}}{4n} \dfrac{9}{8} F^4 \int_0^\infty \dfrac{d\ell}{\ell} \ell^5\\
&\int d\m^{(4)}_{123|4} \left[\dfrac{\sin \p \b_1 \cos \p \b_3}{\sin \p (\b_1+\b_2) \cos \p (\b_2+\b_3)} \right]^{-\ap s} \left[\dfrac{\sin \p \b_2 \cos \p \b_4}{\sin \p (\b_1+\b_2) \cos \p (\b_2+\b_3)} \right]^{-\ap t}
\end{split}
\end{equation}

\section{Supersymmetry vs Hybrid formalism}
\label{sect:SUSY_vs_hybrid}

As we have checked that the 4-point amplitudes in the RNS formalism are MHV, thus satisfying the expected SUSY Ward identities, we can super-symmetrize them. 

In the $\cN =1$ case, the relevant super-invariant is $W^2\bar{W}^2$ where $W_\alpha={1/2}\bar{D}^2D_\alpha U$ is the linearised super-field strength of the vector multiplet described by a real scalar super-field $U$. The full $\cN =1$ super-amplitude would read
\begin{equation}
\cA^{\cN =1}_\textup{4-pt} = \int d^2\vartheta d^2\bar\vartheta \{(W_1W_2)(\bar{W}_3\bar{W}_4) \cA^{Bose}_{4-pt} + {\rm perms} \}
\end{equation}
In addition to 4 vector boson amplitudes $\cA({-},{-},{+},{+})$, it encodes also 2 vector 2 gluino amplitudes 
$\cA({-1/2},{-},{+1/2},{+})$ and 4 gluino amplitudes $\cA({-1/2},{+1/2},{-1/2},{+1/2})$. 
A direct computation of the one-loop amplitude with 2 or 4 gluini looks quite laborious since in addition to the vertex operator in the `canonical' -1/2 super-ghost picture
\begin{equation}
V^{(-1/2)}_F = u^\alpha(k) S_\alpha \Sigma e^{-\varphi/2} e^{ikX}
\end{equation}
where $\Sigma$ is the internal spin field with R-charge +3/2, one should use also the vertex in the higher +1/2 picture
\begin{equation}
V^{(+1/2)}_F = u^\alpha(k) \partial X_\mu \sigma^\mu_{\alpha\dot\alpha}C^{\dot\alpha} \Sigma e^{+\varphi/2} e^{ikX} + \ldots
\end{equation}

In the $\cN =2$ case, the relevant super-invariant is again $\calW^2\bar{\calW}^2$ where now 
$\calW$ is the chiral super-field describing $\cN =2$ vector multiplets. The full $\cN =2$ super-amplitude would read
\begin{equation}
\cA^{\cN =2}_\textup{4-pt} = \int d^4\vartheta d^4\bar\vartheta \{(\calW_1 \calW_2)(\bar{\calW}_3\bar{\calW}_4) \cA^{Bose}_{4-pt} + {\rm perms}\}
\end{equation}
In addition to 4 vector boson amplitudes, it encodes also 2 vector 2 gluino amplitudes $\cA({-}\frac{1}{2},{-},{+}\frac{1}{2},{+})$, 4 gluino amplitudes $\cA({-1/2},{+1/2},{-1/2},{+1/2})$,  2 vector 2 scalar amplitudes $\cA({-},{+},{0},{0})$, 2 gluino 2 scalar amplitudes $\cA({-}1/2,{+}1/2,{0},{0})$ and 4 scalar amplitudes $\cA(0,0,{0},{0})$.
Once again a direct computation of the one-loop amplitudes with gluini looks more involved, while amplitudes with scalars look feasible, since the vertex operator for scalars in vector multiplets are simply  
\begin{equation}
V_\phi = \phi(k) (i\partial Z_3 + k{\cdot}\psi \Psi_3)e^{ikX}
\end{equation}
where $Z_3$ is the complex `untwisted' coordinate and $\Psi_3$ its world-sheet super-partner. 

Different manifestly supersymmetric formalisms for the quantisation of the superstring have been proposed, depending on the number of space-time and internal dimensions \cite{Berkovits:1994wr, Berkovits:1999im, Berkovits:1999in, Berkovits:2001tg}. The one we will focus on here is suitable for compactifications on Calabi-Yau spaces or orbifolds, described by internal $\cN =2$ SCFT's \cite{Berkovits:1994wr}. In the Type II case these yield $\cN=2$ supersymmetry in $D=4$. In the Type I or Heterotic case these yield $\cN=1$ supersymmetry in $D=4$.  Another one is suitable for compactifications on manifolds with $SU(2)$ holonomy (\eg $K3\times T2$) spaces or orbifolds of $T^4$, described by internal $\cN =4$ SCFT's \cite{Berkovits:1999im}. In the Type II case these yield $\cN=4$ supersymmetry in $D=4$. In the Type I or Heterotic case these yield $\cN=2$ supersymmetry in $D=4$. We will neither deal with the second approach any further here, nor with the pure spinor formalism \cite{Berkovits:2004tw,Bedoya:2009np,Berkovits:2014rpa,Berkovits:2015yra}.

We would like to compare our results with those of the hybrid formalism with minimal supersymmetry in $D=4$\cite{Berkovits:1994wr, Berkovits:2001nv}. The construction is based on two observations. First one can twist $\cN =2$ SCFT's redefining the worldsheet stress tensor $T$ according to $T' = T-{J/2}$ where $J$ is the $U(1)$ worldsheet current. As a result $c'=0$ and $h'=h-q/2$. Second one can identify 
the dimension 0 spin fields $\vartheta_a = S_a e^{+\varphi/2}$ with the Grassmann coordinates of superspace. The construction works for $D=4$ whereby $\vartheta_a \rightarrow \vartheta_\alpha , \bar\vartheta_{\dot\alpha}$ 
denoting by $p_\alpha , \bar{p}_{\dot\alpha}$ their dimension 1 conjugate momenta, one has 4 $(\eta,\xi)$ systems with $c=-8=4\times(-2)$. Including the $c=4$ contribution of the bosonic coordinates $X^\mu$ and the $c=0$ contribution of the twisted internal $\cN =2$ SCFT, one has a defect $\Delta{c} = -4$ that can be compensated by 
an additional chiral boson $\rho$ with $\varepsilon = 1$ (as for commuting ghosts like $\varphi$ for $\beta,\gamma$), background charge $Q_\rho=1$ (instead of $Q_\varphi = 2$) and central charge $c=\e(1+3Q^2)=+4$ (instead of $c_\varphi = 13$). 
After twisting, the mapping of the generators of the twisted $\cN =2$ SCFT read
\begin{gather}
T_{hyb} {=} T_X {+} T_{\vartheta} {+} T_\rho {+} T_{tSCFT} {=} T_{RNS} {=} T_{X\psi} {+} T_{SCFT} {+} T_{gh} \quad , \quad J_{hyb} {=} -\partial\rho {+} J_{tSCFT} {=} bc {+} \xi\eta\\
G^+_{hyb} = e^\rho d^2 + G^+_{tSCFT} = J_{BRST} \,\, , \,\, G^-_{hyb} = e^{-\rho} \bar{d}^2 + G^-_{tSCFT} = b
\end{gather}
where
\begin{gather}
d_\alpha = p_\alpha +  \frac{i}{2} \partial{X}_{\alpha\dot\alpha}\bar\vartheta^{\dot\alpha} - \frac{1}{4} \bar\vartheta^2 \partial\vartheta_{\alpha}  + \frac{1}{8} \vartheta_{\alpha} \partial(\bar\vartheta^2) \\
\bar{d}_{\dot\alpha} = \bar{p}_{\dot\alpha}+  \frac{i}{2} \vartheta^{\alpha} \partial{X}_{\alpha\dot\alpha}- \frac{1}{4} \vartheta^2 \partial\bar\vartheta_{\dot\alpha}  + \frac{1}{8} \bar\vartheta_{\dot\alpha} \partial(\vartheta^2)
\end{gather}

Unintegrated (`c-ghost-number 1' in a sense) vertex operators  for compactification independent states, such as an open string gauge boson or the closed string graviton, can be expressed in terms of a real scalar super-field $U(x,\vartheta, \bar\vartheta)$  (or $U(x,\vartheta_L, \bar\vartheta_L; \vartheta_R, \bar\vartheta_R)$ for closed strings). In order for $U$ to be a world-sheet super-primary it must satisfy $D^2U = \bar{D}^2U = \partial_\mu\partial^\mu U = 0$.  Component fields obtain by acting with super-derivatives. For the vector boson one has
\begin{equation}
A_{\alpha\dot\alpha} = [D_{\alpha},\bar{D}_{\dot\alpha}] U\vert_{\vartheta=\bar\vartheta=0}
\end{equation}
and for the gaugino one finds
\begin{equation}
\lambda_{\alpha} = \frac{1}{2} \bar{D}^2 D_{\alpha}U\vert_{\vartheta=\bar\vartheta=0}
\end{equation}

Integrated vertex operators follow the twisting prescription
\begin{equation}
\int dz V = \int dz G_{hyb}^-G_{hyb}^+U = \int dz HU 
\end{equation}
where the field-dependent super-differential operator $H$ reads
\begin{equation}
H =  d^{\alpha} \bar{D}^2 D_{\alpha} + \bar{d}_{\dot\alpha} D^2 \bar{D}^{\dot\alpha} + \partial\vartheta^{\alpha}D_{\alpha} + \partial\bar\vartheta_{\dot\alpha}\bar{D}^{\dot\alpha} + \frac{i}{2} \Pi^{\dot\alpha\alpha}[D_{\alpha},\bar{D}_{\dot\alpha}]
\end{equation}
with $\Pi^{\dot\alpha\alpha} = \partial{X}^{\dot\alpha\alpha} {-} i \bar\vartheta^{\dot\alpha}\partial \vartheta^\alpha {+} i \partial\bar\vartheta^{\dot\alpha} \vartheta^\alpha$. The formalism is manifestly supersymmetric since $d^{\alpha}$, $\bar{d}_{\dot\alpha}$ and $\Pi^{\dot\alpha\alpha}$ commute with the space-time supersymmetry generators $\cQ_{\alpha}$ and $\bar{\cQ}_{\dot\alpha}$. Gauge invariance corresponds to $\delta U = \bar{D}^2\bar\Lambda + {D}^2\Lambda = \Omega + \bar\Omega$ that adds a total derivative to $V$. 

The `topological/twisted' prescription for computing one-loop Type II scattering amplitudes \cite{Berkovits:2001nv} reads
\begin{equation}
\cM_n = \int {d^2\tau\over \tau_2^2} \int \prod_{i=1}^n d^2z_i \langle 
(\int J_L\wedge J_R)^2 V_1 V_2 \ldots V_n \rangle
\end{equation}
where $\int J_L\wedge J_R = \int d^2w (J_L^{tSCFT} - \partial\rho_L)(J_R^{tSCFT} - \partial\rho_R)$, constructed from the twisted $U(1)$ current, is needed to provide the correct number of zero-modes in the large Hilbert space once translated into RNS fields. 

For Type I amplitudes the analogous prescription would be
\begin{equation}
\cA_n = \int {dT\over T} \int \prod_{i=1}^n dz_i \int dw_1dw_2 \langle 
 J(w_1) J(w_2) V_1 V_2 \ldots V_n \rangle
\end{equation}

As observed in \cite{Berkovits:2001nv} when the external states, such as gauge bosons or gravitons, are $\rho$ and compactification independent, the internal contribution factorizes and one has to simply compute contractions of the space-time super-coordinate fields. 

Indeed after dealing with subtleties associated to the integration over the chiral boson $\rho$ one arrives at a manifestly supersymmetric result,
\begin{equation}
\begin{split}
\cM_n{=}& {\int} d^2\vartheta_L d^2\bar\vartheta_L d^2\vartheta_R d^2\bar\vartheta_R {\int} {d^2\tau\over \tau_2^6} {\int} \prod_{i=1}^n d^2z_i {\partial\over d\zeta_i}{\partial\over d\bar\zeta_i} \!\!\left[ (\sum\zeta_iW^i_L)^2(\sum\zeta_i \bar{W}^i_L)^2 (\sum\bar\zeta_iW^i_R)^2(\sum\bar\zeta_i  \bar{W}^i_R)^2 \right. \\
& \times \left. |{:}e^{\cS}{:}|^2 \exp(-2\pi[\cK + i \sum_i z_i k_i]^2/\tau_2)  \prod_{i<j} |\theta_1(z_{ij})|^{k_i{\cdot}k_j} \prod_i \tilde{U}(k_i,\vartheta_L, \bar\vartheta_L, \vartheta_R, \bar\vartheta_R)\right]_{\zeta_i=\bar\zeta_i=0}
\end{split}
\end{equation}
where $\zeta_i$ and $\bar\zeta_i$ are auxiliary Grassmann variables that serve the purpose to select the multi-linear term in the `external polarisations'
\begin{equation}
\cK^\mu = \sum_i \zeta_i B^\mu_i - i \sigma^\mu_{\alpha\dot\alpha} \sum_{i,j} S_{ij} \zeta_i\zeta_j W_i^{\alpha}
\bar{W}_j^{\dot\alpha}
\end{equation}
and 
\begin{equation}
\cS = \sum_{i,j} \zeta_i S_{ij} (D^{\alpha}_jW^i_{\alpha} + \bar{D}^{\dot\alpha}_j\bar{W}^i_{\dot\alpha} - i k_j^\mu B_\mu^i) + \cO(\zeta^2) + \cO(\zeta^3) + \cO(\zeta^4)
\end{equation}
whose consistency has been tested at least for the uncompactified case against gauge invariance, modular invariance and periodicity and equivalence with the RNS formalism. 
 
 The above formula drastically simplifies for $n=4$, since all the derivatives with respect to $\zeta_i$ and $\bar\zeta_i$ must act on the explicit factor not on the exponents and produce $W_L^2\bar{W}_L^2 W_R^2\bar{W}_R^2$. One can safely set $\cK=0$ and $\cS=0$ and get the expected result, \ie the BGS formula for scattering of Type II super-gravitons in $D=10$ written in a notation suitable for $\cN =2$ $D=4$, \ie for `compactification independent' states:   $\cN =2$ supergavity $\{ g_{\mu\nu}, 2\psi_\mu, A_\mu\}$ and $\cN =2$ dilaton hyper-multiplet $\{ \varphi, b_{\mu\nu}, 2\zeta, c_{\mu\nu}, \alpha \}$. These are precisely the states that one gets combining two $\cN=1$ vector multiplets, one for the Left- and one for the Right-movers. 
 
Taking the `square root' of the closed-string result one gets the Type I superstring super-amplitude
\begin{gather}
\cA^\textup{super}_n = \int d^2\vartheta d^2\bar\vartheta  \int {dT\over T^6} \int_\cR \prod_{i=1}^n dz_i {\partial\over d\zeta_i}\\
\left[(\Sigma_i\zeta_iW^i)^2(\Sigma_i\zeta_i \bar{W}^i)^2{:}e^{\cS}{:} e^{-{2\pi \over T} [i \cK +\Sigma_i z_i k_i]^2}  \prod_{i<j} |\theta_1(z_{ij})|^{k_i{\cdot}k_j} \prod_i \tilde{U}(k_i,\vartheta, \bar\vartheta)  \right]_{\zeta_i=0}
\end{gather}
Focussing on the $n=4$ case, one can safely set $\cK$ and $\cS$ to zero and get
\begin{equation}
\cA^{super}_n = \int d^2\vartheta d^2\bar\vartheta  W^2 \bar{W}^2 \int {dT\over T^6} \int_\cR \prod_{i=1}^n dz_i e^{-{2\pi \over T} [\Sigma_i z_i k_i]^2}  \prod_{i<j} |\theta_1(z_{ij})|^{k_i{\cdot}k_j}
\end{equation}
This precisely coincides with our result for the $\cN = 4$ sector in the decompactification limit,  where $\Lambda \approx 1/T^3$ (in the absence of an IR regulator) and indeed the internal SCFT is free and decouples from the space-time part.  

Since for $\cN = 1,2$ sectors, the internal contribution does not simply factorize, even when the external states, such as gauge bosons or gravitons, are $\rho$ and compactification independent, but rather produce derivatives of the Witten index, we expect the hybrid approach to fail to give the correct result if not properly amended. The success at tree level is largely due to the fact that tree-level amplitudes for gluons or gravitons are independent of the amount of supersymmetry (in so far as only minimal couplings are present) and the super-symmetrization is unique when the number of supersymmetry is chosen / given.  

The probable source of the disagreement with the $D=4$ hybrid formalism \cite{Berkovits:2001nv} are the subtleties in defining the functional integration over the chiral boson $\rho$\footnote{We thank Nathan Berkovits for suggesting this interpretation.}.  The difference between the $\cN=1,2$ and the $\cN=4$ contributions to 1-loop amplitudes is that different numbers of fermionic zero-modes can come from the compactification-dependent part of the world-sheet action through the term 
$R_{mnpq} \Psi_L^m \Psi_L^n \Psi_R^p \Psi_R^q$ where $R_{mnpq}$ is the CY curvature which couples to the left and right-moving internal fermions $\Psi_L^m$ and $\Psi_R^n$. In the $\cN=4$  sector, there is a cancellation (see Eq.(3.2) in \cite{Berkovits:2001nv}) between the functional integral over the $\rho$ field and over fermionic zero modes coming from the compactification. Probably this cancellation does not occur in the $\cN=1$ and $\cN=2$ sectors and the coupling of $\Psi_L^m$ and $\Psi_R^n$ through the CY curvature affects the factorization properties.

\section{Conclusions}
\label{sect:conclusions}

We have shown that the 4-point vector boson amplitudes computed in \cite{Bianchi:2006nf} satisfy the correct supersymmetry Ward identities in that they vanish for non MHV helicity configurations $({+}{+}{+}{+})$ and $({-}{+}{+}{+})$. In the MHV case $({+}{+}{-}{-})$ we have simplified their expressions to an extremely compact form. The integrands only involve 
three functions $\cE_\cN$, $\cC_{\cN=1}$ and $\cF_\cN$ of the relevant modular parameter $T$, of the brane configuration $u^I_{ab}$ and the `compactification' moduli, the ubiquitous Koba-Nielsen factor $\Pi(z_i,k_i)$ and the non-holomorphic function $\cY(z_{ij}) = -2 [\cP(z_{ij}) - S^2(z_{ij})]$ with no poles in $z_{ij}$. In $\cN=4$ sectors  $\cE_{\cN=4} = 0$ and only $\cF_{\cN =4} \approx \Lambda_\parallel$ plays a role. Somewhat unexpectedly we have found that no massless poles in two-particle channels are exposed in the $\cN=1,2$ sectors either, thanks to the regular behaviour of $\calY$.

We have then studied the limiting IR and UV behaviour, confirming standard expectations. 

Relying on the supersymmetric properties of the result we have generalised our bosonic amplitudes to manifestly supersymmetric super-amplitudes and compared the results with those obtained in the hybrid formalism and found it hard to reconcile the contributions of  ${\cN =1,2}$ sectors that can be ascribed to subtleties in performing the functional integral of the chiral boson $\rho$. We hope one could find a way to overcome this problem and reproduce  the results we found in the RNS formalism within the (minimal) hybrid formalism. Alternatively, one could address the same issues within the pure spinor approach \cite{Berkovits:2004tw,Bedoya:2009np,Mafra:2011nv,Mafra:2011nw,Berkovits:2014rpa,Berkovits:2015yra} if one could find a reliable way to partially break supersymmetry.  

Based on our present analysis, there are various directions that one can explore: higher number of insertion points, higher loops, more realistic brane configurations or closed string amplitudes. Let's comment on  these extensions. 

Concerning higher points, 5-points look feasible since only MHV or anti-MHV amplitudes should be non vanishing, 6-points looks harder since also NMHV amplitudes corresponding to $(+++---)$ helicity configurations  should be non-vanishing. Factorizations in three- and higher-particle channels could be analysed and the soft behaviour in string theory could be studied more systematically, extending the tree level analyses \cite{Bianchi:2014gla,Bianchi:2015yta,Bern:2014vva,DiVecchia:2015bfa,DiVecchia:2015oba,Ademollo:1975pf}.

Some two-loop results are accessible and at three loops there is some work \cite{Bianchi:1988ux,Bianchi:1989du} but starting at four loops one should expect conceptual problems in addition to practical ones \cite{Donagi:2013dua,Sen:2015hia}.

Barring some subtleties, it should be almost straightforward to generalise our manifestly supersymmetric results to closed superstring amplitudes and explicitly check if any form of KLT relations may be hidden in the connection.

Last but not least, phenomenologically more appealing configurations than `regular' branes should be easy to address \cite{Angelantonj:2002ct,Blumenhagen:2006ci,Lust:2008qc,Anchordoqui:2009mm,Lust:2009pz,Uranga:2012string}.

\vskip 1cm

\section*{Acknowledgments}
Discussions with A.~Addazi, M.~Berg, N.~Berkovits, A.~Guerrieri, C.~Mafra, J.~F.~Morales, O.~Schlotterer, C-K.~Wen are kindly acknowledged. This work was partially supported by the ERC Advanced Grant n.226455 {\it ``Superfields''}, by INFN ST\&FI {\it ``String Theory and Fundamental Interactions''}
and by the Uncovering Excellence Grant STAI {\it ``String Theory and Inflation''}.  

\appendix

\section{Spinor helicity formalism}
\label{sect:spinor_helicity_formalism}

In $D=4$ one can introduce helicity spinors $u_\a^i$ and $\bar{u}_{\dot\a}^i$ such that $k_{i \alpha\dot\alpha} = u_\a^i\bar{u}_{\dot\alpha}^i$ and $a^{i+}_{\alpha\dot\alpha} = v_\alpha^i\bar{u}_{\dot\alpha}^i/ \sqrt{2} v^{i\beta}u_\beta^i$ as well as
$a^{i-}_{\alpha\dot\alpha}= u_\alpha^i\bar{v}_{\dot\alpha}^i/ \sqrt{2} \bar{u}^{i\beta}\bar{v}_\beta^i$, with $v_\alpha^i$ and $\bar{v}_{\dot\alpha}^i$ arbitrary Weyl spinors associated to gauge variations in that $v_\alpha^i \rightarrow v_\alpha^i + \l u_\alpha^i$ yields 
\begin{equation}
a^{i+}_{\alpha\dot\alpha} \rightarrow  a^{i+}_{\alpha\dot\alpha} + \l {u_\alpha^i \bar{u}_{\dot\alpha}^i\over v^{i\beta}u^i_\beta} = a^{i+}_{\alpha\dot\alpha} + {\l  \over v^{i\beta}u^i_\beta} k^i_{\alpha\dot\alpha}
\end{equation}
and similarly for $a^{i-}_{\alpha\dot\alpha}$. The bilinears can be written as $u_i^\alpha u_{j\alpha} = \langle i j \rangle = -  \langle j i \rangle$, $\bar{u}_{i\dot\alpha} \bar{u}_{j}^{\dot\alpha} = [ i j ] = - [j i]$ and 
$2k_i{{\cdot}}k_j =  - \langle i j \rangle [ i j ]$, momentum conservation $\sum_i |i\rangle[i| = 0 = \sum_i |i] \langle i|$. For more details see {\it e.~g.} \cite{Elvang:2013cua}.

A fundamental ingredient in our analysis are the traces of the tensors $f$'s representing the linearised field-strengths. We introduce a compact notation: $(f_1 \dots f_n)= \tensor{f}{^{\mu_1}_{\mu_2}}\dots \tensor{f}{^{\mu_n}_{\mu_1}}$. This quantities are gauge and Lorentz invariant by definition. In our computation we meet traces with two, three and four $f$'s, we need to calculate their values for fixed helicity configurations. 
It is also useful to decompose the tensors $f$ in positive and negative helicity part $ f_{\m \n}= f_i^+ \bar{\S}_{\m \n}^i+f_{i'}^-\S_{\m \n}^{i'}$ with $i,i'=1,2,3$. The matrices $\S$ and $\bar{\S}$ provide the bases of the representations $3_L$ and $3_R$ of $SL(2,\setC)$ thus they are self-dual and antiself-dual matrices and satisfy 
\begin{gather}
\tensor{\levi}{_\m _\n ^\r ^\s} \bar{\S}_{\r \s}^i=+i\bar{\S}_{\m \n}^i \quad
\tensor{\levi}{_\m _\n ^\r ^\s} \S_{\r \s}^{i'}=-i\S_{\m \n}^{i'} \quad
\bar{\S}^i \S^{i'}=\S^{i'} \bar{\S}^i \qquad \S^{i'}\S^{j'}= \d^{i' j'} \idnty + i \tensor{\levi}{^{i'}^{j'}_{k'}} \S^{k'}\\
\Tr  \S^{i'} = \Tr  \bar{\S}^i = 0 \quad \Tr \S^{i'}\bar{\S}^i=0 \quad
\Tr \S^{i'}\S^{j'}=4 \d^{i' j'} \quad  \Tr \bar{\S}^i \bar{\S}^j=4 \d^{i j}
\end{gather}
The product of two vectors in $(1,0)$ and $(0,1)$ can be related to traces in the Lorentz indices: $4 f^+ \!\! {\cdot} g^+ =4 f^+_i (g^+)^i =  (f^+ g^+) $ and $4 f^- \!\!{\cdot} g^- =(f^- g^-)$.
Lorentz invariance helps to recognize when a trace vanishes: self-dual and antiself-dual matrices cannot contract with one another. In particular traces with an odd number of $f$'s with positive (or negative) helicity vanish.

We start to compute traces from products of two $f$'s. We have two independent cases: $({-}{+})$ and $({+}{+})$. $(f_1^- f_2^+)=0$ is zero for Lorentz invariance. In the second case we can use the gauge choice $q_1=q_2=q$ to cancel some terms and obtain
\begin{equation}
(f_1^+ f_2^+)=2 a_1^+{\cdot} k_2 \, a_2^+ {\cdot} k_1 =\dfrac{[12]\<2q\>}{\<1q\>}\dfrac{[21]\<1q\>}{\<2q\>}=-[12]^2
\end{equation}
For traces with three $f$'s, we have two independent cases: $({-}{+}{+})$ and $({+}{+}{+})$. $(f_1^- f_2^+ f_3^+)=0$ is zero for Lorentz invariance. For $({+}{+}{+})$ we use the gauge $q_1=q_2=q_3=q$
\begin{equation}
(f_1^+ f_2^+ f_3^+)=a_1^+{\cdot} k_3 \, a_2^+{\cdot} k_1 \, a_3^+{\cdot} k_2 - a_1^+{\cdot} k_2 \, a_2^+{\cdot} k_3 \, a_3^+{\cdot} k_1 =-\dfrac{1}{\sqrt{2}} [12][23][31]
\end{equation}
For traces with four $f$'s, we have three independent cases: $({+}{+}{+}{+})$, $({-}{+}{+}{+})$ and $({-}{-}{+}{+})$. $(f_1^- f_2^+ f_3^+ f_4^+){=}0$ is zero for Lorentz invariance. For $({+}{+}{+}{+})$ we use the gauge $q_1{=}q_2{=}q_3{=}q_4{=}q$
\begin{equation}
(f_1^+ f_2^+ f_3^+ f_4^+)=a_1^+{\cdot} k_4 \, a_2^+{\cdot} k_1 \, a_3^+{\cdot} k_2 \, a_4^+{\cdot} k_3 + a_1^+{\cdot} k_2 \, a_2^+{\cdot} k_3 \, a_3^+{\cdot} k_4 \, a_4^+{\cdot} k_1=\dfrac{1}{2} [12][23][34][41]
\end{equation}
To compute $({-}{-}{+}{+})$ we choose the gauge $q_1=q_2=k_3$ and $q_3=q_4=k_2$
\begin{equation}
(f_1^- f_2^- f_3^+ f_4^+)= a_1^-{\cdot} a_4^+ \, a_2^-{\cdot} k_1 \,a_3^+{\cdot} k_4 \, k_2 {\cdot}k_3= \dfrac{1}{4} \<12\>^2 [34]^2 = -\dfrac{1}{4} \dfrac{\<12\>^3}{\<23\>\<34\>\<41\>} s t
\end{equation}

\section{Elliptic functions}
\label{sect:ellipric_functions}

Let $q=e^{2\p i \t}$, with $\t$ complex, the Jacobi $\theta$ functions are defined as 
\begin{equation}
\q [\substack{\a \\ \b}] (z|\t)=\sum_{k\in \setZ} q^{(k-\a)^2/2} e^{2\p i(z-\b)(k-\a)}
\end{equation}
with $\a$ and $\b$ real numbers, representing the spin structures. $\theta$ functions solve the heat equation $4\p i \der_\t \q = \der^2_z \q$. There are identifications between different values of $\a$ and $\b$
\begin{equation}
\q [\substack{\a+k \\ \b}] (z|\t)= \q [\substack{\a \\ \b}] (z|\t) \qquad
\q [\substack{\a \\ -\b}] (z|\t)= \q [\substack{-\a \\ \b}] (-z|\t) \qquad
\q [\substack{\a \\ \b+k}] (z|\t)= \q [\substack{\a \\ \b}] (z-k|\t)
\end{equation}
This functions enjoy pseudo-periodicity properties
\begin{equation}
\q [\substack{\a \\ \b}] (z+k|\t)= e^{-2 \p i k \a} \q [\substack{\a \\ \b}] (z|\t) \quad
\q [\substack{\a \\ \b}] (z+k\t|\t)= e^{-2\p i k (z-\b +\t/2)} \q [\substack{\a \\ \b}] (z|\t)
\end{equation}
Under the modular transformations $T$ and $S$, one finds
\begin{gather}
\q [\substack{\a \\ \b}] (z|\t+k)= e^{-i \p k \a(\a-1)} \q [\substack{\a \\ \b+k(\a-1/2)}] (z|\t) \\
\q [\substack{\a \\ \b}] \bigg(\frac{z}{\t} \bigg| -\frac{1}{\t}\bigg)= (-i\t)^{1/2} e^{i \p (2\a \b +z^2/\t)} \q [\substack{\b \\ -\a}] (z|\t)
\end{gather}
Often we omit the dependence on $\t$. For our purposes, we are interested in four particular $\theta$ functions
\begin{equation}
\begin{split}
\q [\substack{1/2 \\ 1/2}] (z|\t)&=\q_1 (z|\t)= 2q^{1/8} \sin (\p z) \prod_{n=1}^\infty (1-q^n) (1-e^{2\p i z} q^n) (1-e^{-2\p i z} q^n) \\
\q [\substack{1/2 \\ 0}] (z|\t)&=\q_2 (z|\t)= 2q^{1/8} \cos (\p z) \prod_{n=1}^\infty (1-q^n) (1+e^{2\p i z} q^n) (1+e^{-2\p i z} q^n) \\
\q [\substack{0 \\ 0}] (z|\t)&=\q_3 (z|\t)= \prod_{n=1}^\infty (1-q^n) (1+e^{2\p i z} q^{n-1/2}) (1+e^{-2\p i z} q^{n-1/2}) \\
\q [\substack{0 \\ 1/2}] (z|\t)&=\q_4 (z|\t)= \prod_{n=1}^\infty (1-q^n) (1-e^{2\p i z} q^{n-1/2}) (1-e^{-2\p i z} q^{n-1/2})
\end{split}
\end{equation}
Under $z \leftrightarrow -z$ $\q_1$ is odd and $\q_2$, $\q_3$ and $\q_4$ are even.

There are other two other ubiquitous elliptic functions: the Dedekind function $\h(\t)=q^{1/24} \prod_{n=1}^\infty (1-q^n)$ and the Weierstrass function $\calP(z,\t)=\der_z^2 \log \q_1 (z|\t)-2 \h_1(\t)$, where $\h_1(\t)=-2\pi i \der_\t \log \h(\t)$. Dedekind function is related also to $\q_1^\prime$ by $\q_1^\prime(0)= 2 \pi \h^3$. Under modular transformations, Weierstrass function is a modular form with weight two and Dedekind function transforms as
\begin{equation}
\calP\left(\dfrac{z}{c\t+d},\dfrac{a\t+b}{c\t+d}\right)=(c\t+d)^2\calP(z,\t) \quad \h(\t+1)=e^{i\p/12} \h(\t) \quad \h \left(-\frac{1}{\t}\right)=(-i\t)^{1/2} \h(\t)
\end{equation}
The contractions of free bosons and fermions at one-loop are:
\begin{gather}
\dbraket{X_{\m_1} \, X_{\m_2}}=\ap \h_{\m_1 \m_2} \calG_\S(z_{12}) \quad
\dbraket{\y_{\m_1} \, \y_{\m_2}}_\a=\ap \h_{\m_1 \m_2} S_\a(z_{12}) \\
\dbraket{\der X^{\m_1} \, e^{i k_2 X_2}}= i \ap k^{\m_1} \der_1 \calG_\S(z_{12}) \quad
\dbraket{\der X^{\m_1} \, \der X^{\m_2}}= \ap \h^{\m_1 \m_2} \der_1 \der_2 \calG_\S(z_{12})
\end{gather}
Where $\calG_\S$ is the bosonic propagator (Bargmann kernel) on the surface and $S_\a$ is the fermion propagator (Szego kernel). The bosonic propagator on the torus is
\begin{equation}
\calG_T(z_1,z_2;\t)=-\dfrac{1}{2} \left[\log \left| \dfrac{\q_1(z_1-z_2|\t)}{\q_1^{\prime} (0|\t)} \right|^2-2 \pi \dfrac{[\im(z_1-z_2)]^2}{\im \t}\right]
\end{equation}
It's easy to see that $\calG_T(z_1,z_2;\t)$ is even under $z_1 \leftrightarrow z_2$, bi-periodic and quasi-invariant under modular transformation: the unique non-trivial transformation is $\calG_T (z_1/\t,z_2/\t;-1/\t)= \calG_T(z_1,z_2;\t)+ \half \log |\t|^2 $.
We can say that the propagator is an ``inhomogeneous" modular form of degree zero. The annulus obtains from the torus by means of the involution $\tilde{z}=1-\bar{z}$ and the modular parameter $\t_\calA=iT/2$ \cite{Sagnotti:1987tw, Bianchi:1988fr, Pradisi:1988xd, Bianchi:1988ux, Bianchi:1989du}, so the propagator between two points $z_1$ and $z_2$ is the sum of the propagators on the torus between $z_1$ and all the images of $z_2$, i. e. $z_2$ itself and $1-\bar{z}_2$.
\begin{equation}
\calG_\calA(z_1,z_2;\t_\calA)=\dfrac{1}{2}\left[\calG_T(z_1,z_2;\t_\calA)+\calG_T(z_1,1-\bar{z}_2;\t_\calA)+\calG_T(1-\bar{z}_1,z_2;\t_\calA)+\calG_T(1-\bar{z}_1,1-\bar{z}_2;\t_\calA) \right]
\end{equation}
On the annulus boundaries ($z=\t \n$ or $z=\t \n+1/2$), the annulus propagator takes the form
\begin{equation}
\label{eq:annulus_propagator}
\calG_\calA(z_1,z_2;\t_\calA)=-2 \left[\log\dfrac{\q_1 (z_1-z_2|\t)}{\q_1^\prime (0|\t)}-\p \dfrac{[\im(z_1-z_2)]^2}{\im \t} \right]
\end{equation}
The M\"obius strip can be obtained in may ways, one consists in using the same involution used to construct the annulus but considering a torus with modular parameter $\t_\calM=\t_\calA+1/2$ \cite{Sagnotti:1987tw, Bianchi:1988fr, Pradisi:1988xd, Bianchi:1988ux, Bianchi:1989du}. The propagator is similar to the annulus propagator
\begin{equation}
\calG_\calM(z_1,z_2;\t_\calM)=\calG_\calA(z_1,z_2;\t_\calM)
\end{equation}
The explicit formula of the Szego kernel or fermionic propagator is
\begin{gather}
S_\a(z_1,z_2;\t)=
\begin{cases}
-\der_1 \calG_\calA(z_1,z_2), & \text{if $\a=1$,} \\
\dfrac{\q_\a(z_1-z_2)}{\q_1(z_1-z_2)} \dfrac{\q_1^\prime(0)}{\q_\a  (0)} , & \text{if $\a \neq 1$.}
\end{cases}
\end{gather}
where $\calG_\calA$ is not computed on the boundaries. $S_\a$ is an odd function under $z_1 \leftrightarrow z_2$. We often use $S_{ij}$ instead $S_1(z_{ij})=-\der_i \calG_\calA(z_{ij})$. The fermionic propagator is a bi-periodic modular form with weight one. When the insertions are on a boundary of the annulus, the explicit form of $S_{ij}$ is
\begin{equation}
S_{ij}=-\pder{\,}{z_i} \left(\calG_\calA (z_i,z_j) \big|_{\bar{z}=-z} \right)=
2 \left[\dfrac{\q_1^\prime (z_{ij})}{\q_1 (z_{ij})}-2 \p i \dfrac{\im z_{ij}}{\im \t_\calA} \right]
\end{equation}
For non-planar amplitude it is necessary to compute a propagator between two points on different boundaries. The propagator can be obtained with the parametrization $z=\t \n +1/2$
\begin{equation}
\calG^T_\calA(z_1{,}z_2){=}\calG_\calA(\t \n_1{,}\t \n_2{+}1/2){=}{-}2\!\left[\log\dfrac{\q_2 (\t \n_{12}|\t)}{\q_1^\prime (0|\t)}{-}\p \dfrac{\im^2 z_{12}}{\im \t} \right] \,\, S_{ij}^T=\!2 \!\left[\dfrac{\q_2^\prime (\t \n_{ij})}{\q_2 (\t \n_{ij})}{-}2 \p i \dfrac{\im z_{ij}}{\im \t_\calA} \right]
\end{equation}

In the computation we bear in mind some properties of $\theta$ functions and propagators. In the even spin structures, products of fermionic propagators can be simplified using the two identities
\begin{equation}
S_\a^2(z,\t)=\calP(z,\t)-e_{\a-1}(\t) \quad,\quad S_\a(z_{12})S_\a(z_{23})=-S_\a(z_{13}) \w_{123}-S^\prime_\a(z_{13})
\end{equation}
where $e_{\a-1}(\t)=4\pi i \der_\t \ln [\q_{\a-1}(0|\t)/\h (\t)]$ and $\w_{123}=S_{12}+S_{23}+S_{31}$. Deriving wrt $z$ the function $\calP-S_\a^2$ we find $2 S_\a(z) S_\a^\prime (z)= \der_z \calP(z)$, thus the derivative of $S_\a^2$ is independent of $\a$.

There is a formula that links the Weierstrass function and the square of the propagator derivative, it's linked to a function that we call $\calY(z)$
\begin{equation}
\calY(z)= -2 \left[\calP(z)-S^2(z) \right]
\end{equation}
where $S(z)$ is in its general form, not computed on a boundary. This function is a modular form with weight two and has no poles. One can prove a generalized Fay identity \cite{Fay:fay,Broedel:2014vla}
\begin{equation}
\W_{123}=S_{12}S_{23}+S_{23}S_{31}+S_{31}S_{12}=-\calY_{12}-\calY_{23}-\calY_{31}
\end{equation}
In order to study the limiting behaviour of $\calY$ it is convenient to first consider the expansion of $\q_1$ in powers of $z$
\begin{equation}
\q_1(z)=z\q_1^\prime(0) \left[1-\h_1 z^2+\dfrac{1}{5}\left(\h_2+3\h_1^2\right)z^4 \right] +\calO(z^7)
\end{equation}
We use this expansions to compute $\calY(z)$
\begin{equation}
\begin{split}
\calY(z=iT\n/2)&=-\dfrac{1}{2}\left[-\der_z^2 \log \q_1 (z)-2\h_1 - \left(\der_z \q_1 (z)-2\p i \dfrac{\im z}{\im \t} \right)^2 \right] = \\
&=-8\left(\h_1+\dfrac{2\p}{T}\right)-8 \left(\dfrac{T^2}{10}(\h_2+3\h_1^2)+\p T \h_1+\p^2\right)\n^2+ \calO(\n^4)
\end{split}
\end{equation}
See \cite{Broedel:2014vla} for more details on relations between elliptic functions and string one-loop amplitudes.

\subsection{Vanishing contractions due to Riemann identities}

Amplitudes assume the form $ \sum_\alpha c_\alpha A_\alpha \partition$, with $A_\alpha$ some function of $z$ and $\t$. Contractions with one bilinear are zero because of normal ordering, $\<{:}\y_1^{\m_1} \y_2^{\m_1}{:}\>_\a=0$. In the even sector contractions of bosonic operators only give zero. To see that we use the Riemann identity:
\begin{equation}
\label{eq:riemann_identity}
\sum_{\alpha\neq 1}c_\alpha \theta_\alpha(z_1)\theta_\alpha(z_2)\theta_\alpha(z_3)\theta_\alpha(z_4) 
= \theta_1 (z_1')\theta_1 (z_2')\theta_1 (z_3')\theta_1 (z_4')- \theta_1 (z_1'')\theta_1 (z_2'')\theta_1 (z_3'')\theta_1 (z_4'')
\end{equation}
Where the variables $z_i'$ and $z_i''$ are linear combinations of the $z_i$
\begin{equation}
\begin{pmatrix}
z_1' \\ z_2' \\ z_3' \\ z_4'
\end{pmatrix}= \dfrac{1}{2}
\begin{pmatrix}
1 & 1 & 1 & 1 \\
1 & 1 & -1 & -1 \\
1 & -1 & 1 & -1 \\
1 & -1 & -1 & 1
\end{pmatrix}
\begin{pmatrix}
z_1 \\ z_2 \\ z_3 \\ z_4
\end{pmatrix} \qquad
\begin{pmatrix}
z_1'' \\ z_2'' \\ z_3'' \\ z_4''
\end{pmatrix}= \dfrac{1}{2}
\begin{pmatrix}
-1 & 1 & 1 & 1 \\
1 & -1 & 1 & 1 \\
1 & 1 & -1 & 1 \\
1 & 1 & 1 & -1
\end{pmatrix}
\begin{pmatrix}
z_1 \\ z_2 \\ z_3 \\ z_4
\end{pmatrix}
\end{equation}
The $\theta_1(z)$ functions vanish when $z=0\mod 1$, thus this quantity is zero if at least one of the $z_i'$ and one of the $z_i''$ are zero. Considering a generic purely bosonic correlator $\braket{{\cal O}_1 \dots {\cal O}_n}_\alpha= \braket{\braket{{\cal O}_1 \dots {\cal O}_n}} \partition $, the sum over the even spin structures produces a vanishing result in the most general case with $\calN=1$ supersymmetry, in that
\begin{equation}
\sum_{\a \neq 1} c_\alpha \partition \propto \sum_{\a \neq 1} c_\alpha \theta_\alpha(0) \theta_\alpha (u_{ab}^1) \theta_\alpha (u_{ab}^2) \theta_\alpha (u_{ab}^3)=0
\end{equation}
To understand why this expression is zero it is enough to consider the first line of the two matrices: $2 z_1'=2 z_1''=u_{ab}^1+u_{ab}^2+u_{ab}^3=0 \mod 1$.

\printbibliography
\end{document}